\documentclass[twoside,11pt]{article}

\usepackage[preprint]{jmlr2e}

\usepackage[utf8]{inputenc}
\usepackage{url}
\usepackage{microtype}
\usepackage{graphicx}
\usepackage{subfigure}
\usepackage{booktabs} % for professional tables
\usepackage{algorithm}
\usepackage{algorithmic}
\usepackage{amsmath,amsfonts,amssymb} %% amsthm already in jmlr2e.sty
\usepackage{bbm}
\usepackage{color}
\usepackage{mathrsfs}
\usepackage{bm}
\usepackage{multirow}
\usepackage{booktabs}
\usepackage{makecell}
\usepackage{graphicx}
\usepackage{subfigure}
\usepackage{caption}
\usepackage{thmtools}
\usepackage{times}
\usepackage{cancel}
\usepackage{url}
\usepackage{bm}
\usepackage{booktabs}
\usepackage{lineno}
\usepackage{enumitem}

\usepackage[utf8]{inputenc} % allow utf-8 input
\usepackage[T1]{fontenc}    % use 8-bit T1 fonts
\usepackage{hyperref}       % hyperlinks
\usepackage{url}            % simple URL typesetting
\usepackage{booktabs}       % professional-quality tables
\usepackage{amsfonts}       % blackboard math symbols
\usepackage{nicefrac}       % compact symbols for 1/2, etc.
\usepackage{microtype}      % microtypography
\usepackage{xcolor}         % colors

\newcommand{\MSE}{\widehat{\textsc{MSE}}}

\newcommand{\MoM}{\text{MoM}}
\newcommand{\score}{\hat{\textbf{S}}}

\usepackage[capitalize]{cleveref}
\crefname{nlem}{Lemma}{Lemmas}
\crefname{nprop}{Proposition}{Propositions}
\crefname{ncor}{Corollary}{Corollaries}
\crefname{nthm}{Theorem}{Theorems}
\crefname{exa}{Example}{Examples}
\crefname{assumption}{Assumption}{Assumptions}
\crefname{equation}{}{}

\newcommand{\mat}[1]{\ensuremath{\mathbf{#1}}}

\newcommand{\Ind}[1]{\mathbbm{1}{#1}}

\newcommand{\abs}[1]{|{#1}|}

\newcommand{\mE}{\mathbb{E}}

\newcommand{\A}{\mat{A}}
\newcommand{\B}{\mat{B}}

\newcommand{\X}{\mat{X}}

\newcommand{\cC}{\mathcal{C}}

\newcommand{\cT}{\mathcal{T}}

\newcommand{\cX}{\mathcal{X}}
\newcommand{\cY}{\mathcal{Y}}

\usepackage{lastpage}
\jmlrheading{23}{2022}{1-\pageref{LastPage}}{1/21; Revised 5/22}{9/22}{21-0000}{Author One and Author Two}

\ShortHeadings{Tripuraneni, Perrault-Joncas, Madeka, Foster and Jordan}{Meta-Analysis of Randomized Experiments}
\firstpageno{1}

\begin{document}

	\title{Meta-Analysis of Randomized Experiments with Applications to Heavy-Tailed Response Data}

	\author{%
		\name Nilesh Tripuraneni\thanks{Work done while at Amazon.} \\
		\addr University of California, Berkeley\\
		\AND
		\name Dominique Perrault-Joncas \thanks{Correspondence to joncas [at] amazon dot com.} \\
		\addr Amazon, Seattle  \\
		% examples of more authors
		\AND
		\name Dhruv Madeka \\
		\addr Amazon, NYC\\
		\AND
		\name Dean Foster\\
		\addr Amazon, NYC \\
		\AND
		\name Michael I. Jordan \\
		\addr University of California, Berkeley, Amazon \\
		%   \And 
		%   Nan Jiang\\
		%   University of Illinois, Urbana-Champaign
		%   \And 
		%   Dhruv Madeka\\
		%   Amazon, NYC
	}

	\editor{}
	
	\maketitle
	
	\begin{abstract}%   <- trailing '%' for backward compatibility of .sty file
			A central obstacle in the objective assessment of treatment effect (TE) estimators in randomized control trials (RCTs) is the lack of ground truth (or validation set) to test their performance. In this paper, we propose a novel cross-validation-like methodology to address this challenge. The key insight of our procedure is that the noisy (but unbiased) difference-of-means estimate can be used as a ground truth ``label" on a portion of the RCT, to test the performance of an estimator trained on the other portion. We combine this insight with an aggregation scheme, which borrows statistical strength across a large collection of RCTs, to present an end-to-end methodology for judging an estimator's ability to recover the underlying treatment effect as well as produce an optimal treatment "roll out" policy. We evaluate our methodology across 699 RCTs implemented in the Amazon supply chain. In this heavy-tailed setting, our methodology suggests that procedures that aggressively downweight or truncate large values, while introducing bias, lower the variance enough to ensure that the treatment effect is more accurately estimated.
		
	\end{abstract}
	
	%\begin{keywords}
	%	keyword one, keyword two, keyword three
	%\end{keywords}

	%!TEX root = main.tex
\section{Introduction}
Causal inference is widely used across numerous disciplines such as medicine, technology, and economics to inform important downstream decisions \citep{hernan2020}. Inferring causal relationships between an intervention and outcome requires estimating the
treatment effect (TE): the difference between what happened given an intervention and what would have happened in its absence. A central difficulty is that these two events are never jointly observed \citep{rubin2005causal}. TE estimation leverages randomized controlled trials (RCTs)---which randomly assign the items of interest into either the treatment or control groups---to counter selection biases and allow causal effects to be estimated via a simple differences-in-means estimate.

Indeed, the simplest ``model-free" unbiased estimator of a treatment effect is the difference-in-means (DM) estimate \citep{rubin2005causal}. Such an estimator may, however, suffer from high variance in real-world scenarios which often involve  heterogeneous, high-dimensional and heavy-tailed data\footnote{Such heavy-tailed data is commonplace in the large-scale RCTs which motivate our study.}. A plethora of additional information is thus often used to improve TE estimates relative to this simple baseline. For example, pretreatment regression adjustments can significantly reduce the variance of a treatment effect estimate while adding little additional bias \citep{angrist2008mostly, imbens2015causal}. Similarly, a host of other regularization and robustness modifications can be used to trade off bias and variance.

As the complexity of such estimators increases, so do the assumptions (and work) needed to establish their statistical validity. One particular setting in which this becomes easier, and which we argue arises in many practical applications,\footnote{Including AB testing of forecasting model improvements, website changes, supply-chain modifications, or a number of other interventions.} % do we subset everythin to AB tests? There are probably quite a few labs with mulitple ARMS. I get that claim 1 is based on AB tests but it generalizes just as well for other RCTs.
is when large RCTs can be run on the same population. This setting provides an opportunity to get at the fundamental attributes of interest---the mean-squared error (MSE) of a given treatment effect estimator and its ability to inform treatment roll out decisions. Our simple insight is that the DM estimator can function as a noisy, but unbiased ``label" for the treatment effect. Noisy estimates for a TE estimator performance can then be computed by comparing this estimator to the (unbiased) difference-in-means estimator via a simple, held-out validation estimate (see \cref{claim:1} and \cref{claim:2}). Our goal in this work is to judge the performance of TE estimators by pooling noisy (but unbiased) estimates of their performance \textit{across many RCTs}. % This is basically a variation on empirical risk minimization (with a nice twist where the target comes from a secondary iid sample). We should call a spade a spade I think.
%is when large RCTs can be run in the same population. This setting provides an opportunity to get at the fundamental quantity of interest---the mean-squared error (MSE) of a given treatment effect estimator. Our simple insight is that the DM estimator can function as a noisy, but unbiased ``label" for the treatment effect. Noisy MSE estimates for a TE estimator can then be computed by comparing this estimator to the (unbiased) difference-in-means estimator via a simple, held-out validation estimate (see \cref{claim:1}). Our goal in this work is to judge the performance of TE estimators by pooling many noisy (but unbiased) estimates of their MSEs \textit{across many RCTs}. % This is basically a variation on empirical risk minimization (with a nice twist where the target comes from a secondary iid sample). We should call a spade a spade I think.
Such a procedure is desirable because it targets the actual quantity of interest, the estimator MSE, in an assumption/estimator-agnostic fashion. The primary contributions of this work are as follows:
\begin{itemize}[leftmargin=.5cm]
	\item We process a corpus of 699 genuine RCTs implemented at Amazon across several years and we highlight the heavy-tailed nature of the response and covariate variables. The unique challenges associated with heavy-tailed estimation require careful navigation of the bias-variance tradeoff which motivates the development of an objective selection procedure for TE estimation.
	\item We present a selection scheme which borrows statistical strength across the corpus of RCTs in order to judge the relative performance of several commonly used TE estimators, including their usefulness at defining a treatment roll out policy.
	\item We use this framework to argue that in the presence of heavy-tailed data---that often arise in large-scale technology and logistics applications---aggressive downweighting and truncation procedures are needed to control variance.
	\item We propose an extension of this methodology that allows us to use a collection of RCTs to assess the impact of different roll out policies for an RCT.
	\item We also use this framework to show that the generally accepted practice to use statistical significance at level $\alpha=0.05$ for the TE to determine the roll out policy is far from optimal for the Amazon Supply Chain and should instead be determined empirically.
\end{itemize}

\subsection{Related Work}
\label{sec:related}
The literature on causal inference and treatment effect estimation is vast and a comprehensive review is beyond the scope of this paper. \cite{hernan2020, imbens2015causal, angrist2008mostly, hadad2020} and \cite{wager2020stats} provide modern perspectives on both the theory and practice of treatment effect estimation. Cross-validation (CV) also has been (and remains) a major subject of statistical inquiry as it is amongst the most widely used tools to assess the quality of an estimator and perform model selection \cite{bayle2020crossvalidation, lei2020cross, stone1974cross, geisser1975predictive}.

Relatively little work has been done in the intersection of these two domains. Part of the difficulty stems from the fact that the standard procedure of CV breaks down for treatment effect estimation since the true treatment effect is never observed in  data. \cite{athey2016recursive} and \cite{powers2018some} do provide model-specific selection methods in the context of treatment effect estimation. However, these works do not apply to arbitrary TE estimators. Closest to our work is that of \cite{schuler2018comparison}, who use a data-splitting methodology to evaluate several risk functions to assess  \textit{heterogeneous} treatment effect estimators. This differs from our work in two principal ways. First, our framework is targets the problem of \emph{average} treatment effect estimation---in many scenarios that we are interested in, treatments cannot be individualized and must be applied in an all-or-nothing fashion to the entire population. Our statistical scheme also differs since we provide a provably \textit{unbiased} estimate\footnote{Leveraging the unbiased nature of the DM estimator.} of the mean-squared error of a TE estimator, % Provably unbiased follows from empirical risk formulation. We should be explicit about that. I know the proof is trivial but as stated it seems like it's something we should prove. If we can attribute the proof the empirical risk then no proof is needed I think.
and we introduce an aggregation scheme to borrow statistical strength across different AB tests to compare estimators. Additionally, our work uses a large corpus of 699 \textit{actual} randomized AB tests conducted at Amazon over the course of several years as our test-bed for estimator selection in contrast to synthetic data simulations.

One of our main motivations is to highlight the unique challenges associated with heavy-tailed data often present in applications arising at large-scale technology and logistics companies. Semiparametric TE estimators for heavy-tailed datasets inspired by similar applications have been explored \cite{fithian2014semiparametric} and \cite{taddy2016scalable}. However, these works do not address the problem of model selection which is our central focus. Specifically, we focus on methods to select among simple estimators (with few to no tuning parameters) that are widely used in practice. % Technically we restrict our selves to simple estimator because we only have 699 data points. We probably want to be upfront about it. We do this to avoid overfitting giving the fact that we can't easily increase this dataset.

\subsection{Preliminaries}
\label{sec:prelim}
%\textbf{Notation}:
%We use bold-faced variables such as $\X$ and $\x$ to define vectors.

We work within the Rubin potential outcomes model \citep{rubin2005causal} where we imagine we are given a domain of objects $\cY$ and a target variable of interest $Y(\cdot)$ given a possible intervention. For a fixed intervention $I$, our goal is to estimate the population average treatment effect (ATE):
\begin{align}
	\Delta = \mE[Y(1)-Y(0)], \label{eqn:trueATE}
\end{align}
where $Y(1)$ corresponds to the value of an experimental unit---in our case a product in the supply chain---given the treatment and $Y(0)$ its unobserved counterfactual control (and vice versa). In general, we also allow the existence of other covariates in our model $\X \in \cX$. In a given AB test, we first randomly sample an equal number of items into a treatment group, $\cT$, and a control group $\cC$. We further let the ($\X_i, T_i, Y_i$) be the covariates, treatment dummy, and value of the $i$th item. By a standard argument, using the assumption of randomization (independence of $\{ Y_i(1), Y_i(0) \}$ and $T_i$), the differences-in-means estimator,
\begin{align}
	\hat{\Delta}_{DM}= \frac{1}{\abs{\cT}} \sum_{i \in \cT} Y_i(1) - \frac{1}{\abs{\cC}} \sum_{i \in \cC} Y_i(0), \label{eq:dm}
\end{align}
provides an unbiased estimate of $\Delta$ \citep{rubin2005causal}. A primary benefit of the DM estimator is that it is ``model-free." That is, it makes no explicit assumptions on the data-generation process for $Y_i$ as a function of the other covariates.

\subsection{Dataset Description}
\label{sec:data}
We use 699 RCTs that were run at Amazon since 2017 on a population of products. The interventions in each RCT consist of various modifications and (potential) improvements to the way in which products are processed through the supply chain. The RCTs are most often constructed with 50$\%$ of products in an RCT randomly placed in the treatment group and 50$\%$ in the control group, though some are not evenly balanced. The RCTs vary in size from tens of thousands of products to those with several millions. Each RCT is run over the course of approximately 27 weeks with the intervention instituted at a trigger date at 10 weeks in the treatment group. 
% Check; most labs last 8-12 weeks, not 27. Are we including the pre-period in this? Seems misleading.

At each week in an RCT, the response variable generated from each product is computed. Each RCT was preprocessed to contain the averaged pretreatment response (denoted $X$), a strictly nonnegative averaged pretreatment auxiliary covariate (denoted $D$), averaged posttreatment response (denoted $Y$), and binary treatment indicator (denoted $T$) for each item. Auxiliary covariates (such as $D$) often arise in naturally occurring applications where it is feasible to forecast a related quantity to $Y$ (such as the number of expected products needed in a time period to satisfy user demand). 

% In the Appendix we perform an investigation into a single RCT to demonstrate that the response distribution over the range of product has a \textit{heavy tail} -- and in fact behaves as a power law $\sim y^{-\eta}$ with exponent between $1-3$. Estimation in this setting is difficult and requires balancing several considerations when considering the pros and cons of various estimation techniques. Our exploration of these issues serves a dual purpose: (1) to highlight the ubiquitous occurrence of such heavy tails in naturally occurring data, and (2) to motivate the need for a model selection procedure to navigate the bias-variance trade-off.
	%!TEX root = main.tex
\section{Heavy Tails and Hard Estimation Case Study}

The difficulties associated with treatment effect estimation of an intervention in large-scale commerce RCT datasets are many fold.
% "an" intervention? do we assume that all RTCs are really the same intervention over and over?
The most salient difficulty for our consideration is that the response distribution over the range of products has a \textit{heavy tail}. Similar heavy-tailed distributions are known to exist in user revenue distributions as well as user engagement metrics at large-scale technology companies \citep{fithian2014semiparametric, taddy2016scalable}.
%That is, in the scope of Amazon products, there are a few ASINs which sell in large numbers but many ASINs of which only a few are sold. This often results in situations where a few ASINs generate millions of dollars of ltfcf in a week but millions of ASINs may only generate a few dollars of ltfcf.
Estimation in this setting is difficult and requires balancing several considerations when considering the pros and cons of various estimation techniques. Our exploration of these issues serves a dual purpose: (1) to highlight the ubiquitous occurrence of such heavy tails in naturally occurring data, and (2) to motivate the need for a model selection procedure to navigate the bias-variance tradeoff.

\begin{figure}[!ht]
	\centering
	\begin{minipage}[b]{0.45\linewidth}
		\centering
		\includegraphics[width=0.9\linewidth]{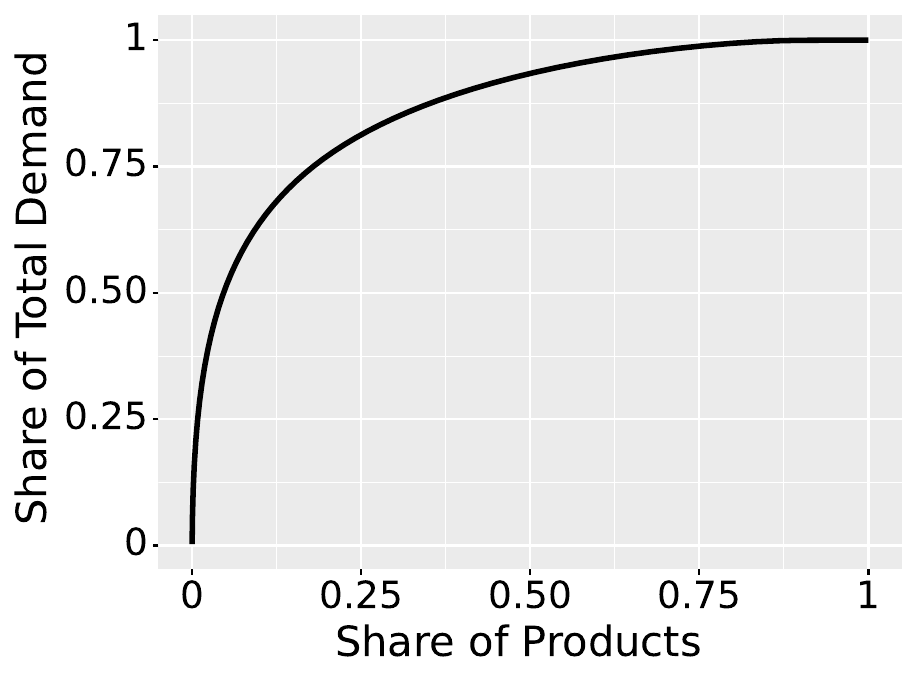}
		\caption{Gini plot of a single RCT showing the cumulative share of demand vs. product population share ordered by descending popularity. Demand is heavy-tailed with the top 20\% most popular products accounting for nearly 80\% of the demand share.}
		\label{fig:dmnd}
	\end{minipage}
	\quad
	\begin{minipage}[b]{0.45\linewidth}
		\centering
		\includegraphics[width=0.9\linewidth]{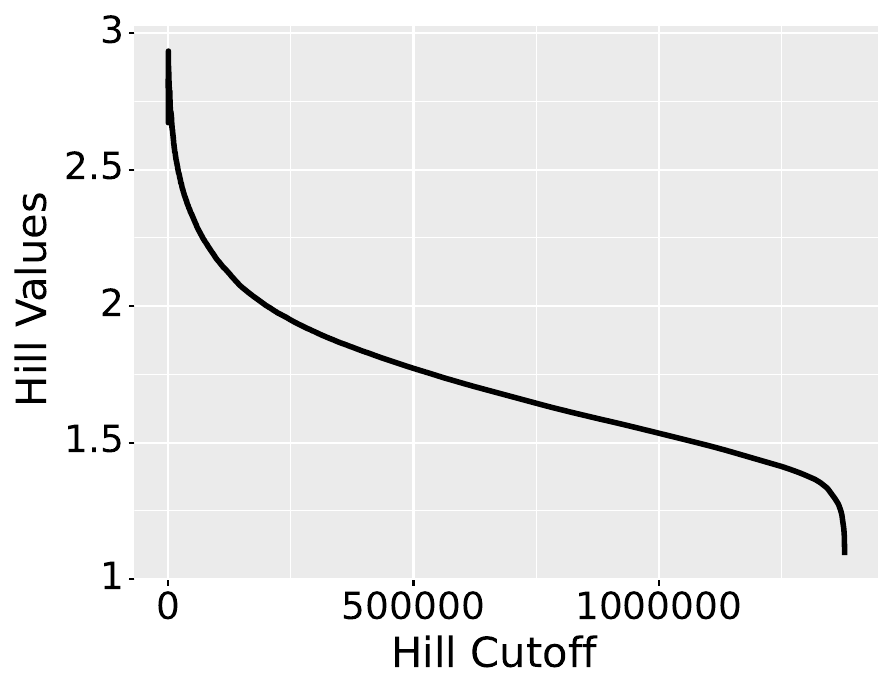}
		\caption{Hill plot of the right tail of the response variable distribution in a single RCT versus the Hill cutoff hyperparameter. The Hill values are an estimate of the power $\eta$ in the asymptotic tail behavior of the response distribution variable, $Y$, $p(y) \sim y^{-\eta}$. }
		\label{fig:hill}
	\end{minipage}
\end{figure}

Let us investigate the data inside a single RCT to assist in further making this point. The RCT under consideration consists of millions of distinct products. This RCT (a representative choice) displays significant heavy-tail behavior, as shown in \cref{fig:hill}.

%\begin{figure}[!ht]
%		\centering
%		\includegraphics[width=0.9\linewidth]{./Figures/demand_share.pdf}
%		\caption{Gini plot in a single RCT showing the cumulative share of demand vs. product population share, with products ordered by descending popularity. Demand is heavy-tailed with the top 20\% most popular products accounting for nearly 80\% of the demand share.}
%		\label{fig:dmnd}
%\end{figure}
%
%\begin{figure}[!ht]
%
%		\centering
%		\includegraphics[width=0.9\linewidth]{./Figures/hill_plot.pdf}
%		\caption{Hill plot of the right tail of the response variable distribution in a single RCT versus the Hill cutoff hyperparameter. The Hill values are an estimate of the power $\eta$ in the asymptotic tail behavior of the response distribution variable, $Y$, $p(y) \sim y^{-\eta}$. }
%		\label{fig:hill}
%
%\end{figure}

%\begin{figure}[!ht]
%\centering
%\begin{minipage}[b]{0.45\linewidth}
%	\centering
%	\includegraphics[width=0.8\linewidth]{one_lab/log_ltfcf.pdf}
%	\caption{50 binned Histogram of the $\sign(ltfcf) \cdot \log(1+ltfcf)$ in the Spock lab. Many products result in no or little cash flow, with a long, heavy right tail of products generating positive cash flow. Note several products also result in negative cash flow, although the right/left tails are highly asymmetric.}
%	\label{fig:ltfcf}
%\end{minipage}
%\quad

% \begin{minipage}[b]{0.9\linewidth}
	% 	\centering
	% 	\includegraphics[width=0.8\linewidth]{./Figures/hill_plot.pdf}
	% 	\caption{Hill plot of the right tail of the response variable distribution in a single RCT versus the Hill cutoff hyperparameter. The Hill values are an estimate of the power $\eta$ in the asymptotic tail behavior of the response distribution variable, $Y$, $p(y) \sim y^{-\eta}$. }
	% 	\label{fig:hill}
	% \end{minipage}
% \end{figure}
%\cref{fig:ltfcf} shows a long, heavy right-tail in the ltfcf distribution.
We implement the Hill estimator to obtain an estimate of the power-law behavior $\eta$ in the right tail distribution of $\sim y^{-\eta}$ across all the RCTs under consideration.
The Hill cutoff hyperparameter is chosen to discard points near the center of the distribution (i.e., near zero) and allows the formulation of a bias-variance tradeoff \citep{drees2000make}. We avoid a more sophisticated data-driven choice of this cutoff since the precise Hill value is not of particular interest in our setting.\footnote{Indeed we have tens of thousands of points in all RCTs, so small-sample difficulties associated with ``Hill horror plots" seem not to arise.}. 
% tens of thousands? there are 7million points in just the RCT chosen for illustration
Rather, it is apparent the power $\eta$ can be conservatively judged to be between $1-3$ in \cref{fig:hill}. Analyzing the response distribution across the entire corpus of 699 RCTs 
% pick AB test or RCT but don't use both expressions simultaneously
and choosing the Hill cutoff parameter at the $5$th percentile shows that the average decay exponent is $\approx 2.32$ with a standard deviation of $0.79$, and median of $2.1476$.

The difficulties seen in this case study reinforce the conclusion that handling the heavy tails inherent in our data likely requires more sophisticated (regularized) estimators than the DM estimator. Ultimately this boils down to balancing the tradeoff between bias and variance in estimation. Navigating this bias-variance tradeoff is one of the primary motivations for our aggregation methodology for TE estimator selection.

\section{Validation Procedure for Treatment Effect Estimators}
\vspace{-0.1cm}
\label{sec:val}
In this section, we present the key idea behind the validation procedure we use to assess the quality of an arbitrary treatment effect estimator, $\hat{\Delta}_{E}(\cdot, \cdot)$, in the RCT denoted $I$. Let $\Delta$ denote the population ATE shown in \cref{eqn:trueATE}. Given the groups $\cT$ and $\cC$, we first randomly partition them into disjoint groups $\cT_1$, $\cT_2$ and $\cC_1$, $\cC_2$. Now, consider the (potentially complicated) treatment effect estimator $\hat{\Delta}_{E}(\cT_1, \cC_1)$ trained on the first fold of data. We can obtain an estimate of its performance by how well it targets the difference-of-means estimator computed on the hold-out set $\hat{\Delta}_{DM}(\cT_2, \cC_2)$:
\begin{align}
	\MSE_{I, E}( (\cT_1, \cC_1), (\cT_2, \cC_2)) = (\hat{\Delta}_{E}(\cT_1, \cC_1) - \hat{\Delta}_{DM}(\cT_2, \cC_2))^2. \label{eq:held_out}
\end{align}
A simple argument shows that this quantity is a \textit{noisy but unbiased} MSE of the estimator (and thus it permits the relative comparison of two different estimators).
\begin{lemma}
	\label{claim:1} % claim is weird, it's expressed in terms of inequality but results is really the equality at the end of the proof. I get that the inequality is enough to order estimators by their MSE, but the point is really that they are all offset by the same DM variance and so empirical risk minimization and estimators ranking works just as usual. 
	Given two different treatment effect estimators $A$ and $B$ in the aforementioned setting, we have:
	\begin{align}
		&  \mE[(\hat{\Delta}_{A}(\cT_1, \cC_1) - \hat{\Delta}_{DM}(\cT_2, \cC_2))^2] \leq \mE[(\hat{\Delta}_{B}(\cT_1, \cC_1) - \hat{\Delta}_{DM}(\cT_2, \cC_2))^2] \implies \label{eq:claim1} \\
		&  \mE[(\hat{\Delta}_{A}(\cT_1, \cC_1) - \Delta)^2] \leq \mE[(\hat{\Delta}_{B}(\cT_1, \cC_1) - \Delta)^2]. \notag
	\end{align}
\end{lemma}
See \cref{appendix:proofs} for a proof. This result motivates using the held-out sample error as a metric to assess the relative merit of two estimators $\hat{\Delta}_{A}$ and $\hat{\Delta}_{B}$. However, simply using this estimator on a single RCT % changed "lab" to RCT; using RCT, lab, RCT to refer to the same thing; pick one and search through the doc for consistency.
provides a (potentially very) noisy estimate of the population error, not the population error itself. Indeed, if the estimator $\hat{\Delta}_{DM}(\cT_2, \cC_2)$ is sufficiently good to estimate $\Delta$, why even bother to use another estimator? Said another way, the error estimate in \cref{eq:held_out} will always suffer at least the variance of the unbiased estimate \cref{eq:dm}. In practice we use a cross-validated version of \cref{eq:held_out} to reduce the subsampling variance due to the random train/test splits (see \cref{appendix:method}). This procedure will not decrease the variance of the DM estimator arising from the underlying heavy-tailed data however.

Our proposal for resolving this conundrum is to note that in many situations we have access to \textit{multiple} RCTs from the same underlying population or process given different interventions. Thus, aggregating the set of error estimates
\begin{align}
	\hat{\A} = \{ \MSE_{I_1, A}((\cT_1, \cC_1), (\cT_2, \cC_2)), \hdots, ...	 \MSE_{I_N, A}((\cT_1, \cC_1), (\cT_2, \cC_2)) \} \label{def:Aset}
\end{align}
and comparing to
\begin{align}
	\hat{\B} =	\{ \MSE_{I_1, B}((\cT_1, \cC_1), (\cT_2, \cC_2)), \hdots, ...	, \MSE_{I_N, B}((\cT_1, \cC_1), (\cT_2, \cC_2)) \}, \label{def:Bset}
\end{align}
for various interventions $\mathcal{I} = \{I_1, \dots, I_N\}$, can allow us to pool information across RCTs.  We sidestep the methodological complexities of performing this aggregation and instead turn to an investigation of simple, practically-motivated schemes.

\subsection{An Aggregation Scheme}
\label{sec:agg}
Aggregating the mean-squared errors requires handling a practical consideration. Since the RCTs and interventions across RCTs themselves may be different, the overall scales of the MSEs between different RCTs may be different. As an example, consider a corpus of two RCTs on which estimator $A$ obtain errors $\{1, 10\}$ and estimator $B$ obtains errors $\{2 , 9\}$. Simply averaging the errors or doing a rank-based test of performance would indicate both estimators are equivalent. However, intuitively we believe a relative improvement of estimator $B$ from $10$ to $9$ on the second RCT does not outweigh the degradation from $1$ to $2$ on the first RCT.

This observation motivates the definition of a normalized score to compare the estimators $A$ vs $B$, as a function of the vectors of their noisy errors.\footnote{As noted earlier, in practice each error estimate is averaged over several resampled train/test splits, but we suppress this extra notation for clarity.} For each intervention $i \in \{I_1,...,I_N\}$ we define the normalized score:
\begin{align}
	S_i(\hat{A}_i, \hat{B}_i) = \frac{\hat{B}_i-\hat{A}_i}{\hat{B}_i + \hat{A}_i}, \label{eq:normalize_obj}
\end{align}
for $\hat{A}_i \in \hat{\A}$ and $\hat{B}_i \in \hat{\B}$. Where $\hat{\A}$ and $\hat{\B}$ are defined according to \cref{def:Aset} and \cref{def:Bset} respectively.

This normalized score vector (which we denote by $\score(\hat{\A}, \hat{\B})$) implicitly binarizes each of its elements to bound them in the range $[-1, 1]$. Each element of this vector is a noisy score of estimator $A$'s performance relative to $B$ \textit{on one RCT in the corpus}.\footnote{Our notion of a normalized score vector is element-wise transitive. That is, $\frac{b-a}{a+b} > 0$ and $\frac{c-b}{b+c} > 0$ imply $\frac{c-a}{a+c}>0$.} If the estimator has many elements that are positive, it suggests that estimator $B$ has larger errors than estimator $A$. In this case, we would expect estimator $A$ to be better than estimator $B$.

To formalize this intuition, we use the following heuristic which implicitly treats each RCT equally independent of size. We use a two-sided one-sample $t$-test applied to this normalized score vector to test the null that the ``population mean" of the $\score$ ``distribution" is $0$, i.e., that the performance of estimator $A$ is indistinguishable from the performance of estimator $B$. Overall, this procedure interpolates between two extremes. A purely rank-based test of performance might only count the number of RCTs for which $A$ is better than $B$ irrespective of how much better one is in a particular RCT. Meanwhile, a procedure which only looks at the raw (unnormalized) RCT errors has the property that RCTs with large MSE values for both estimators would drown out signal from RCTs with small MSE values. We stress that the $t$-test heuristic provides a simple way of converting the information contained in $\score(\hat{\A}, \hat{\B})$ to a single number, but we recommend looking at the score histograms for a more complete picture.

%	\nnote{This intuitively seems reasonable. However, the interventions themselves are arbitrary (and even the pre-treatment/control distributions over $Y$ may be changing in each lab from previously applied interventions) so thinking about each intervention as being sampled from a population of interventions may not be a good model...}

%Perhaps the above notation should also be modified to show intervention dependence and other pretreatment covariates and I should think more about the assumptions etc...}

	%!TEX root = main.tex
\section{Validation Procedure for RCT-Driven Decision Making Policies}
\label{sec:decision}
Though estimating the treatment effect helps teams better understand the inner workings of the Amazon supply chain, the fundamental motivation for running an RCT is typically to determine whether a planned change (a new algorithm, system, or intervention) should be rolled out to all treatment units. Practically, this means deciding whether to deploy a new algorithm, system or other intervention to all products that run through the supply chain. The traditional approach of relying on the treatment effect estimate and its statistical significance to make this decision runs into two types of problems:
\begin{enumerate}[leftmargin=0.5cm]
	\item Statistical significance can be difficult to achieve, especially in the presence of heavy-tailed response data.
	\item The traditional threshold for statistical significance at level $\alpha = 0.05$ is arbitrary and emphasizes controlling for type I errors over type II errors. However, businesses are equally concerned about forgoing opportunities to improve efficiency (type II error) as they are about deploying changes that are not actually beneficial (type I error), so there is no strong reason to favor the status quo. 
\end{enumerate}
%When deciding to roll out a treatment in production, the asymmetry between the two types of errors is of particular concern. The unit of measurement for all RCT treatment effects at Amazon is the weekly financial impact per unit; as such, there is no inherent reason to favor the status quo. The overarching goal is to deploy as many improvements to the supply chain and, to borrow from the bandit literature, minimize regret across RCTs.

This is not to say that the optimal roll out policy is to simply deploy any treatment for which the estimated ATE is positive, i.e. $\hat{\Delta}_E > 0$. Rather, the estimated treatment effect, and associated statistics such as standard error, t-statistic, etc., should be used to guide a binary decision of the form $D(\hat{\Delta}_E) = D_E \in \{0, 1\}$ for deploying a treatment in production. The important insight here is that we are not interested in making only one decision in isolation, but a sequence of decisions over time on the basis of multiple RCTs that share a common metric. In this setting, exploiting commonality between treatment effects allows us to pool information across RCTs and improve our decisions.
From this perspective, it is easier to reason about the problem from a Bayesian perspective by considering the ATE $\Delta_i$ for each RCT is drawn from a prior distribution $\Delta_i \sim \mathbb{P}(\Delta)$. The estimated treatment effect is then given by $\mathbb{P}(\hat{\Delta}_{E,i}|\Delta_i)$, and in that sense our goal can be understood as determining $\mathbb{P}(\Delta_i > 0| \hat{\mathbf{\Delta}}_{E})$, where $\hat{\mathbf{\Delta}}_{E} = (\hat{\Delta}_{E, I_1}, \dots, \hat{\Delta}_{E, I_N})$.

Formulating a fully Bayesian perspective to the problem is neither necessary nor our intention in this paper. We took this detour to stress the fact that $\hat{\Delta}_E > 0$ is not a priori a sensible decision rule even if we consider type I and type II errors to be equally important. In our setting, the goal remains to maximize the cumulative financial impact (CFI) across the corpus of RCTs $\mathcal{I}$, and this can be formalized as the objective function: %, we want to maximize the cumulative financial impact (CFI):
\begin{align}
	F_E = \sum_{i\in\mathcal{I}} M_i \Delta_i D_{E,i} \label{eq:cum_ltfcf}
\end{align}
where $M_i$ is the size of the treatment target population for RCT $i$. i.e. the population to which the treatment would be applied if rolled out.

Directly maximizing \cref{eq:cum_ltfcf} for the best decision rule $D_{E,i}$ is not possible, however, since it would require knowledge of the true ATE  $\Delta_i$ for all RCTs. Fortunately, $F_E$ is linear in $\Delta_i$, and we can produce an unbiased estimate of it using an unbiased estimate for the ATE (provided by the DM estimator $\Delta_{DM, i}$).  To ensure independence between the estimated ATE $\Delta_{DM, i}$ and the decision rule $D_{E,i}$ we can again rely on splitting the treatment and control into two groups $\cT_1, \cT_2$ and $\cC_1, \cC_2$. Doing this for every RCT allows us to construct the empirical objective (see \cref{appendix:method} for details)
\begin{align}
	\hat{F}_E = \sum_{i\in\mathcal{I}} M_i \hat{\Delta}_{DM, i}(\cT_2, \cC_2) D_{E,i}(\cT_1, \cC_1)\label{eq:cum_ltfcf_hat}
\end{align}
which is unbiased for $F_E$:
\begin{lemma} \label{claim:2}
	$\hat{F}_E$ is unbiased for $F_E$ in \eqref{eq:cum_ltfcf} for a decision policy based on sub-sampled data $(\cT_1, \cC_1)$.
\end{lemma}
Again seen \cref{appendix:proofs} for the proof.

We constructed $\hat{F}_E$ with the Amazon supply chain in mind, but it generalizes to any context in which the following apply:  
\begin{enumerate}
	\item The target metric is additive across units in the population and treatments applied.
	\item The treatments do not interact with each other\footnote{In our setting, we believe this assumption to be reasonable in for two reasons. First, since most treatments result in treatment effects of small magnitude, any mutual interactions are approximately locally linear. Second, even for potentially large interactions, such interactions are likely to incoherently add and cancel across RCTs, since there is limited coordination between RCTs  and their purpose.}.
	\item The decision made on the basis of each treatment is binary (i.e. apply treatment or do not). 
    \item The RCTs are statistically independent in that each RCT control/treatment group assignment is independently randomized. 
\end{enumerate}
These requirements are satisfied in many settings. Examples range from field experiments in education aiming to increase standardized testing results in schools, to continuous improvements to user interfaces focused on customer engagement, to medical interventions targeting a specific outcome such as reduced cholesterol, amongst others. Indeed, $\hat{F}_E$ is a relevant quantity in any domain where repeated experiments are the common.

%In a sense, it is a variation on the classification error applied to RCTs.

One nuance worth noting given our lengthy discussion of normalization in \cref{sec:agg} above is that in defining $F_E$, we do not normalize; rather, we track the additive effect of each RCT in proportion to the relevant population to which it would be applied, meaning that RCTs with low ATE whose decision would impact a large population will contribute more than those with high ATE but low corresponding population size--which is sensible from the decision-making perspective. %This also means that XXX and the ATE will have different scale, provided they are all expressed in the same units. 

\section{Results}
In this section we detail several simple and commonly used estimators for TE estimation and subsequently compare their relative performance.
\vspace{-0.2cm}

\subsection{Estimators}

For the following estimators, we note that each admits a ``Winsorization" which can be used to trade off bias and variance. To do this, we can simply Winsorize the covariates and targets, $X, D, Y$, \textit{in only the training fold}, to reduce variance. The test folds are always left untrimmed/Winsorized so \cref{claim:1} remains valid. Explicitly we define Winsorization at level $0.001$ to Winsorize the $X, Y$ distributions at $P0.1$, $P99.9$ and the (positive) auxiliary $D$ distribution at $P99.9$.

% \subsubsection{Difference-of-Means (dm)}
The simple \textbf{difference-of-means estimator},
\begin{align}
	\hat{\Delta}_{DM}= \frac{1}{\abs{\cT}} \sum_{i \in \cT} Y_i(1) - \frac{1}{\abs{\cC}} \sum_{i \in \cC} Y_i(0), \label{eq:dm2}
\end{align} as defined before is the first estimator we consider.
%can be interpreted in a regression framework by writing $Y = Y(0)+T \cdot (Y(1)-Y(0))$. This implies we can estimate the ATE for the binary treatment using linear regression of the observed outcomes $Y_i$ on the vector $(1, T_i)$---which is equivalent to computing \cref{eq:dm2}.
% Under the assumption of randomization, taking expectations conditional on the treatment assignment yields:
% \begin{align}
	% \mE[Y | T] = \alpha + T \Delta \text{ where } \alpha = \mE[Y(0)].\label{eq:dm_cond}
	% \end{align}
%In this setting the noise model is heteroscedastic (and depends on $T$).
\label{subsec:dm}
We also consider the \textbf{Difference-of-Median-of-Means (mom) estimator}
%\subsubsection{Difference-of-Median-of-Means (mom)}
%Our definition of this estimator cannot be interpreted in the regression framework strictly speaking, but it is sufficiently similar that we describe it here. The formulation of \cref{eq:dm2} and relationship to \cref{eq:dm_cond} motivates the robust estimation of $\alpha$ and $\Delta$. %Equivalently, we replace the terms in $\frac{1}{\abs{\cT}} \sum_{i \in \cT} Y_i(1)$, $\frac{1}{\abs{\cC}} \sum_{i \in \cC} Y_i(0)$ with the median-of-means estimators for some prespecified block size $B$ to define,
\begin{align}
	\hat{\Delta}_{DMoM} = \MoM(\{ Y_{i}(1) \}_{i=1}^{\abs{\cT}}	, B) - \MoM(\{ Y_{i}(0) \}_{i=1}^{\abs{\cC}}		, B).
\end{align}
%Similarly, $\hat{\alpha} = \MoM(\{ Y_{i}(0) \}_{i=1}^{\abs{\cC}}		, B)$ to complete the regression analogy.
Where $\MoM(\{ Y_{i}(1) \}_{i=1}^{\abs{\cT}}	, B)$ indicates we bucket the data into $B$ blocks, compute the mean in each block, and the median across all the blocked means.
We use mom$1000$ in our experiments to denote the median-of-means estimator chosen with 1000 total blocks.
% \label{subsec:dmom} 
Next we also consider what we refer to as the \textbf{Generalized Difference-in-Differences (gen$\_$dd) estimator} which assumes access to a pretreatment item-specific covariate $X_i$ corresponding to the response value $Y_i$. So, assuming the model,
\begin{align}
	Y = \alpha + T \cdot \Delta + X \cdot \beta + \epsilon,
\end{align}
we can estimate the ATE for a binary treatment by (least-squares) regressing $Y_i$ onto $(1, T_i, X_i)$, where $\epsilon_i$ represents a general conditionally mean-zero noise term (which may depend on $X_i$). If the covariates $X_i$ are strongly correlated with the response value $Y_i$, incorporating them into the regression can significantly reduce the variance.
Finally we consider a reweighted version of the previous estimator we refer to as the \textbf{Weighted Generalized LR (and Generalized Difference-in-Differences) (gen$\_$dd$\_$w1)} estimator.
% \subsubsection{Weighted Generalized LR (and Generalized Difference-in-Differences) (gen$\_$dd$\_$w1)}
% \label{subsec:gen_dd_weight}
% Since all of the above estimators can be written as various forms of linear regression it is also possible to interpret them from the perspective of $M$-estimation as minimizing a sum of residuals defined as $e_i = Y_i-\alpha-\Delta T_i - X_i \beta$. 
That is, we can consider estimation objectives of the form:
% \begin{align}
	% 	\min_{\alpha, \Delta, \bm{\beta}} \sum_{i=1}^{n} \psi(e_i \cdot w_i),
	% \end{align}
% for some sequence of weights $w_i$. The choice we explore is that of simple weighted least-squares. That is, we take $\psi$ to be the squared loss and define the weights as $w = (1+D)^{-\gamma}$ (for $\gamma>0$) for some nonnegative covariate $D$. 
\begin{align}
	\frac{1}{n} \sum_{i=1}^n \frac{1}{(1+D_i)^\gamma} (Y_i - \alpha -\Delta T_i - \beta_i X_i)^2.
\end{align}
to estimate $\alpha$ $\beta$, and most importantly the TE $\Delta$. In practice, the covariate $D$ is taken as an auxiliary covariate, which serves as positive surrogate capturing the shape of the distribution of $Y$. In this case the weighting has the effect of downweighting large values of $Y$ which can be useful to regularize heavy-tailed distributions.

\subsection{Estimator Comparisons}

In this section, we present results obtained from  a corpus of 699 RCTs performed at Amazon over several years as described in \cref{sec:data}. We compare estimators by their out-of-sample MSE computed via the cross-validation procedure described in \cref{sec:val}. 
%In our experiments, we resampled each treatment and control group into 50/50 splits over 100 different train/test replications. Our reported results for the mean-squared error (MSE) of a given estimator in a lab are averaged over all the held-out validation estimates of the MSE computed on these 100 train/test splits. [repetition]

We begin by studying several of the normalized score histograms to facilitate the comparison of our estimators; additional results are provided in \cref{app:additional}. In judging two estimators $A, B$ via their score distribution $\score(\hat{\A}, \hat{\B})$, we note that a left-skewed score distribution indicates $B$ is a better estimator (in terms of its MSE) than $A$.

\begin{figure}[!htb]
\minipage{0.32\textwidth}
\includegraphics[width=\linewidth]{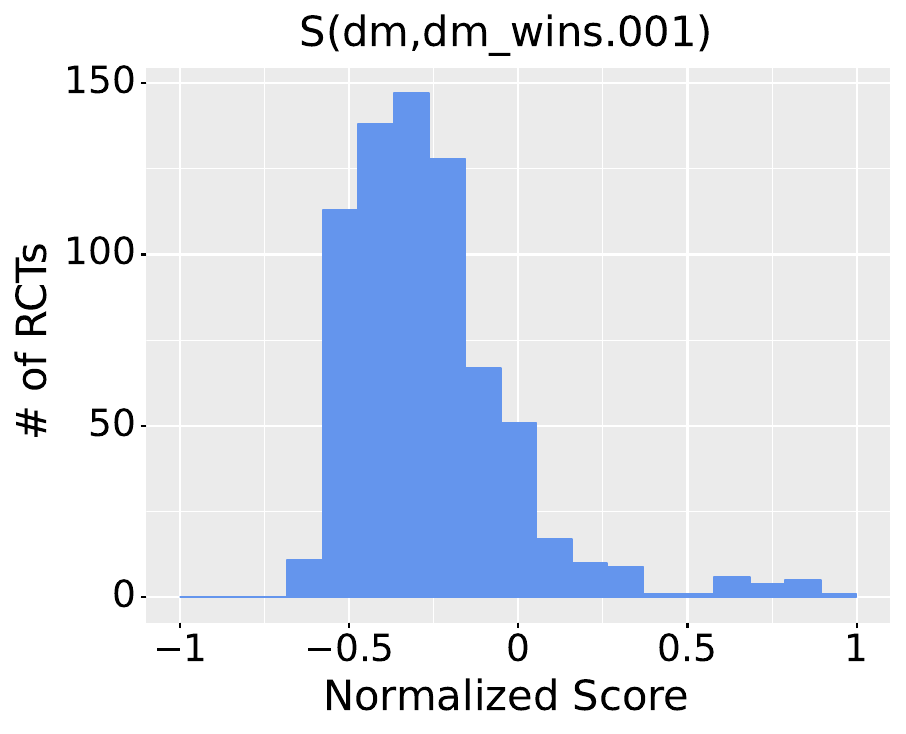}
\caption{Histogram of the score distribution for dm vs Winsorized (at $0.001$) dm estimator.} \label{fig:score_hist_dm_dm_wins}
\endminipage\hfill
\minipage{0.32\textwidth}
\includegraphics[width=\linewidth]{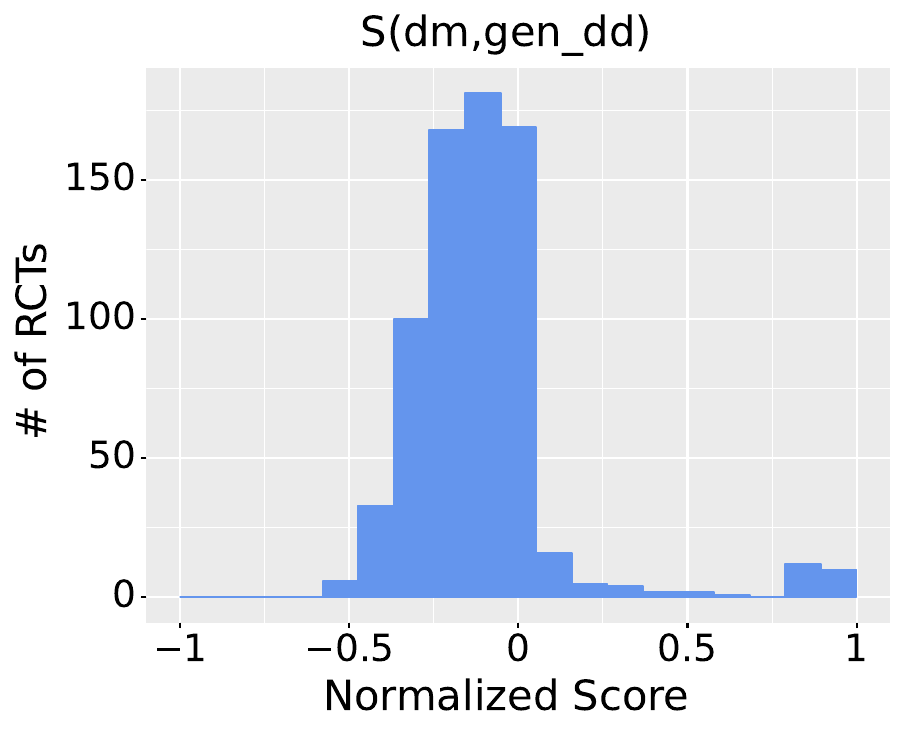}
\caption{Histogram of the score distribution for dm vs gen$\_$dd estimator.} \label{fig:score_hist_dm_gen_dd}
\endminipage\hfill
\minipage{0.32\textwidth}%
\includegraphics[width=\linewidth]{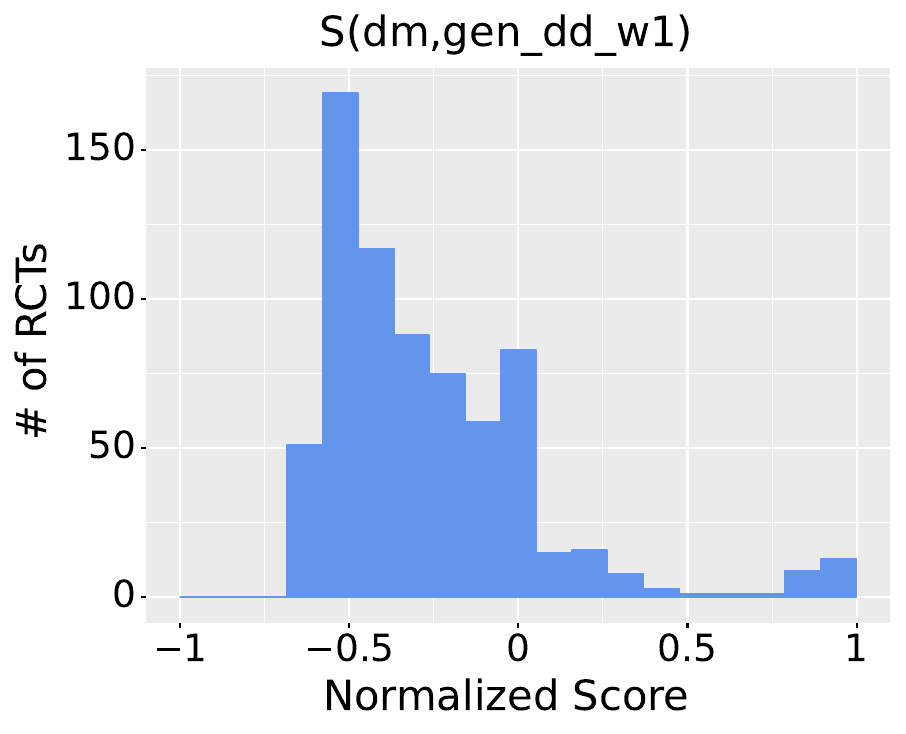}
\caption{Histogram of the score distribution for dm vs gen$\_$dd$\_$w1 estimator.} \label{fig:score_hist_dm_gen_dd_w1}
\endminipage
\end{figure}

In \cref{tab:table0to100}, we use the $t$-test heuristic from \cref{sec:agg} to summarize each score histogram. For the sake of brevity, we do not display all the methods tested in the table.
%We found that all the estimators weighted by different powers ($1,2,3$) of the inverse $D$ distribution perform comparably. 
Overall, we see several phenomena that accord with our expectations. First, adjusting for the pretreatment covariate reduces variance (i.e., gen$\_$dd is better then dm). Second, downweighting large values of $Y$ provides significant value: inverse weighting by $D$ and Winsorization performs generically the best under our metric (gen$\_$dd$\_$w1 and all Winsorized estimators perform well). We also see that the dm estimator is dominated by every other method in \cref{tab:table0to100}; such as the median of median-of-means estimator (mom$1000$), whose robustness underlies its improved performance. 
%We can also check that the results from \cref{tab:table0to100} are stable with respect to different resampled replicates to compute the cross-validated errors which feed into the error vectors in $\hat{\A}$/$\hat{\B}$ as \cref{tab:tab1to50,,tab:tab51to100} in \cref{app:additional} show.

We summarize this table by converting it into a table of pairwise comparisons of wins/losses/ties using a $p$-value to determine the significance of the win or loss. The question of extracting an ordered ranking from the table of wins/losses is a classic problem. The natural procedure of simply summing up the number of row-wise wins is commonly referred to as the Copeland/Borda counting method (see \citep{saari1996copeland} and references within). 
%It has recently been shown that such a simple method is robust to the misspecification of the ranking model and optimal under minimal assumptions \citep{JMLR:v18:16-206}.\footnote{Such consistency results are not entirely necessary since our notion of a normalized score vector is element-wise transitive. That is, $\frac{b-a}{a+b} > 0$ and $\frac{c-b}{b+c} > 0$ imply $\frac{c-a}{a+c}>0$.}

\begin{table}[!htbp]
\caption{Comparison of Estimators via one-sample $t$-test applied to their normalized score vector. Easiest to read row-wise. The index $(A, B)$ of the table computes the pair of the ($t$-statistic, $p$-value) associated with the score $\score(\hat{\A}, \hat{\B})$. A large positive $t$-statistic at index $(A, B)$ indicates estimator $A$ is better then estimator $B$ and vice-versa.}
\label{tab:table0to100}
\centering
\resizebox{\columnwidth}{!}{%
	\begin{tabular}{|l|l|l|l|l|l|l|l|l|}
		\hline
		Method &
		dm &
		mom1000 &
		gen\_dd &
		gen\_dd\_w1 &
		dm\_wins.001 &
		gen\_dd\_wins.001 &
		gen\_dd\_w1\_wins.001 \\ \hline
		dm &
		x &
		(-3.58, 0.000363) &
		(-12.68, 2.38e-33) &
		(-22.36, 3.6e-84) &
		(-28.19, 7.99e-118) &
		(-25.33, 2.96e-101) &
		(-24.96, 4.11e-99) \\ \hline
		mom1000 &
		(3.58, 0.000363) &
		x &
		(-2.12, 0.0342) &
		(-11.89, 7.32e-30) &
		(-13.51, 3.78e-37) &
		(-14.61, 1.94e-42) &
		(-15.72, 5.33e-48) \\ \hline
		gen\_dd &
		(12.68, 2.38e-33) &
		(2.12, 0.0342) &
		x &
		(-21.1, 4.73e-77) &
		(-19.01, 2e-65) &
		(-25.15, 3.11e-100) &
		(-23.49, 1.14e-90) \\ \hline
		gen\_dd\_w1 &
		(22.36, 3.6e-84) &
		(11.89, 7.32e-30) &
		(21.1, 4.73e-77) &
		x &
		(-0.26, 0.794) &
		(-5.12, 3.87e-07) &
		(-9.56, 1.87e-20) \\ \hline
		dm\_wins.001 &
		(28.19, 7.99e-118) &
		(13.51, 3.78e-37) &
		(19.01, 2e-65) &
		(0.26, 0.794) &
		x &
		(-4.17, 3.41e-05) &
		(-5.39, 9.62e-08) \\ \hline
		gen\_dd\_wins.001 &
		(25.33, 2.96e-101) &
		(14.61, 1.94e-42) &
		(25.15, 3.11e-100) &
		(5.12, 3.87e-07) &
		(4.17, 3.41e-05) &
		x &
		(-4.12, 4.2e-05) \\ \hline
		gen\_dd\_w1\_wins.001 & (24.96, 4.11e-99) & (15.72, 5.33e-48) & (23.49, 1.14e-90) & (9.56, 1.87e-20) & (5.39, 9.62e-08) & (4.12, 4.2e-05) & x \\ \hline
	\end{tabular}%
}
\end{table}

Applying such a method by inspection returns the following rankings:
\begin{align}
\hspace*{-0.8cm} \textbf{gen$\_$dd$\_$w1$\_$wins.001} >
\textbf{gen$\_$dd$\_$wins.001} >
\textbf{dm$\_$wins.001} \approx
\textbf{gen$\_$dd$\_$w1} > \textbf{gen$\_$dd} > \textbf{mom1000} > \textbf{dm} \nonumber
\end{align}
Overall, these results suggest that aggressively Winsorizing and/or downweighting heavy tails can profitably trade variance for some additional bias. % We also stress that although our ranking procedure via $t$-statistics is transitive, the score and $t$-statistic values between $A$ and $B$ are computed via relative normalization between just these two estimator errors. Hence the actual values across several estimators are not always directly comparable due to the different normalizations used. Thus, we should always look at the performance of two estimators directly, take their score histogram into consideration, as well as exercise common-sense checks to draw further conclusions.

% the bias is always negative, right (if you remove largest outliers via winsorization)? what's the business cost of that, especially if the known risk of bias isn't communicated? 
	%!TEX root = main.tex
\subsection{Estimator statistic and Roll Out Policies}
\label{res:policy}

We now turn to the question of evaluating the effectiveness of a roll out policy $D_E$. A common policy, which we use internally, is to roll out treatments that are positive and statistically significant at level $\alpha=0.05$. To date, Amazon has used the Generalized Difference-in-Differences estimator, meaning that the standard decision rule has been:  %having a $t$-statistic for the ATE above 1.96\footnote{All of our RCTs have sample size much larger than 30, often in the millions, ensuring that the $t$-statistic and z-statistic are for all purposes equivalent.}. In other words, the standard decision rule used by Amazon is given by: %$D_{\textbf{gen$\_$dd}} = \Ind [\hat{\Delta}_{\textbf{gen$\_$dd}}/\hat{\sigma}_{\textbf{gen$\_$dd}} > 1.96]$.%\frac{\hat{\Delta_{\textbf{gen$\_$dd}}}{\hat{\sigma}} > 1.96]$.
\begin{equation} \label{eq:standard_rule}
	D_{\textbf{standard}} = \Ind{ \left [\frac{\hat{\Delta}_{\textbf{gen$\_$dd}}}{\hat{\sigma}_{\textbf{gen$\_$dd}}} > 1.96 \right ]}
\end{equation}

% An important question then is how does this policy compare to other decision rules using different $t$-critical $t_c$ decision threshold as well as other estimators. To answer this question, \cref{fig:decision_tstat_many} shows the normalized CFI as a function of the $t_c$ for different estimators. That is we plot $\hat{F}_E/\hat{F}_{\textbf{standard}}$ for estimator $E$ as a function of $t_c$ where the decision rule takes the general form:
 
\begin{figure}[!htb]
	\minipage{0.49\textwidth}
	\includegraphics[width=\linewidth]{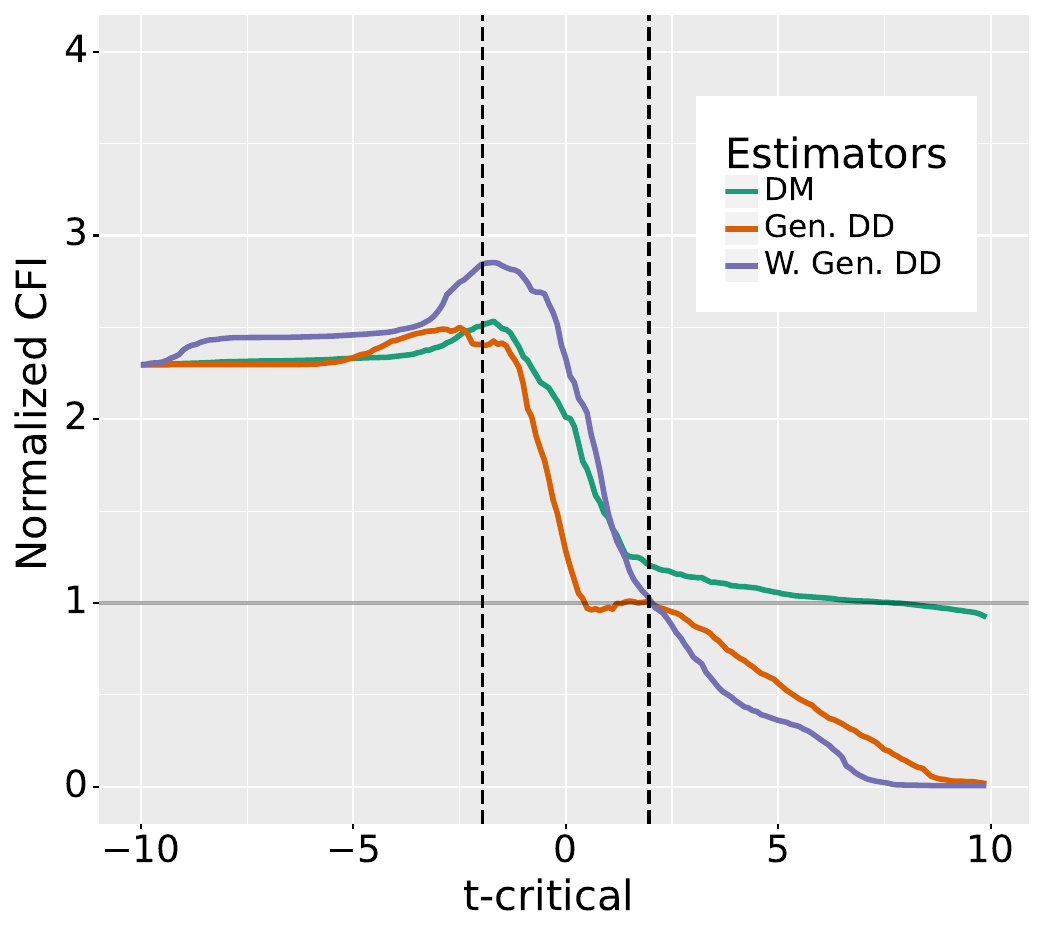}
	\caption{Performance as measured by the normalized CFI for the Difference-in-Means, Generalized Difference-in-Differences, and Weighted Generalized Difference-in-Differences with $\gamma=0.6$ as a function of $t_c$.} \label{fig:decision_tstat_many}
	\endminipage\hfill
	\minipage{0.49\textwidth}
	\includegraphics[width=\linewidth]{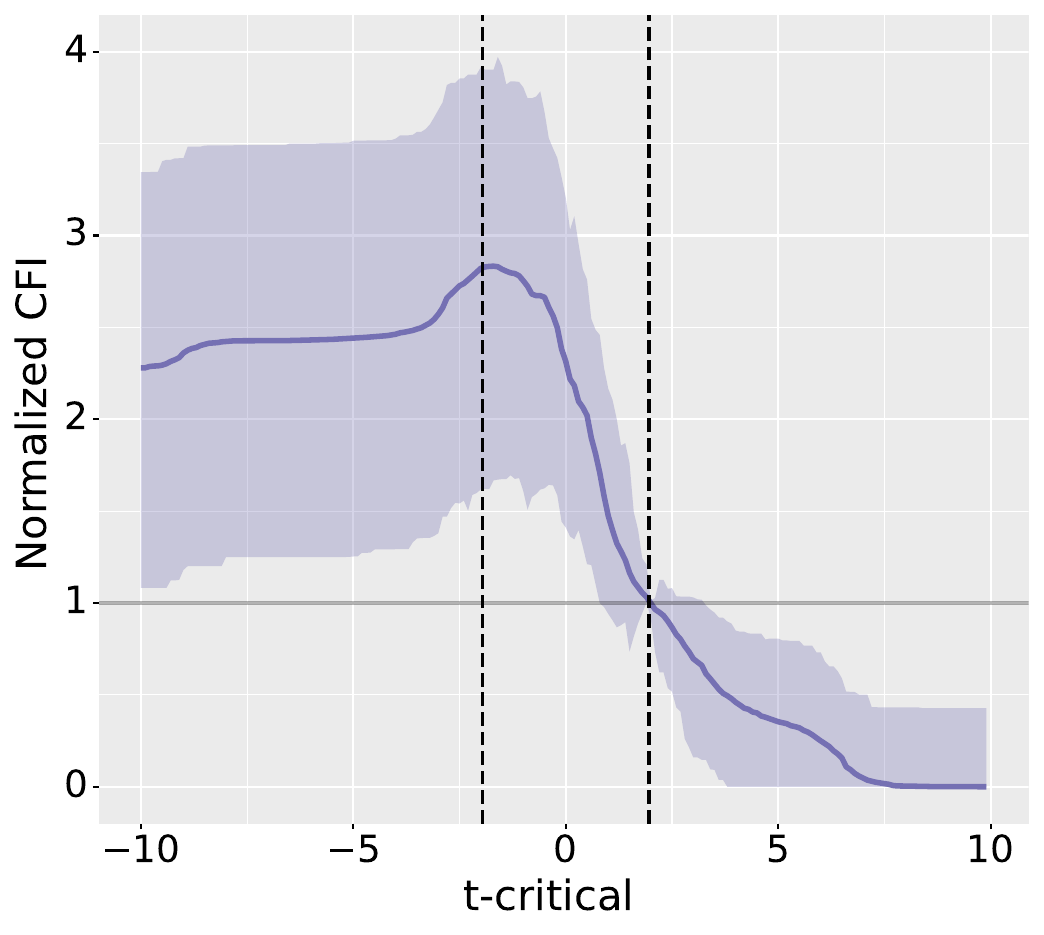}
	\caption{Performance as measured by the normalized CFI for the Weighted Generalized Difference-in-Differences with $\gamma=0.6$ as a function of $t_c$ along with the 95\% confidence bands computed using cross-validation.} \label{fig:decision_tstat}
	\endminipage
\end{figure}

 Because the decision rule depends on the choice of estimator and t-critical $t_c$ decision threshold, it is worth comparing outcomes of various potential combinations, as we do in  \cref{fig:decision_tstat_many}. We plot the normalized CFI $\hat{F}_E/\hat{F}_{\textbf{standard}}$ for estimator $E$ as a function of $t_c$, where the decision rule takes the general form:
\begin{equation} \label{eq:decision_rule}
  D_{E} = \Ind{ \left [\frac{\hat{\Delta}_{E}}{\hat{\sigma}_{E}} > t_c \right ]}.
\end{equation}

 Specifically \cref{fig:decision_tstat_many} shows the normalized CFI, as a function of $t_c$, for 3 estimators:  the Difference-in-Means; the Generalized Difference-in-Differences; and the Weighted Generalized Difference-in-Differences with $\gamma = 0.6$. We included the first two estimators because they correspond to two important baselines: the unbiased ``target'' used to construct $\hat{F}_E$ and the estimator for our standard policy, respectively; we chose the last because it performed best at maximizing $\hat{F}_E$ among the estimators we considered. \cref{fig:decision_tstat} shows the normalized CFI, along with 95\% confidence bands, for the Weighted Generalized Difference-in-Differences with $\gamma = 0.6$. These confidence bands are computed via cross-validation by re-sampling $(\cC_1, \cT_1)$ and $(\cC_2, \cT_2)$ 100 times \footnote{As explained in \cite{bengio2004no}, confidence intervals computed via cross-validation should be interpreted cautiously.}.
 
% \begin{figure}[!htb]
%	\minipage{0.49\textwidth}
%	\includegraphics[width=\linewidth]{Figures/decision_tstat_many_talk.pdf}
%	\caption{Performance as measured by the normalized CFI for the Difference-in-Means, Generalized Difference-in-Differences, and Weighted Generalized Difference-in-Differences with $\gamma=0.6$ as a function of $t_c$.} \label{fig:decision_tstat_many}
%\end{figure}
 %\begin{figure}[!htb]
%	\endminipage\hfill
%	\minipage{0.49\textwidth}
%	\includegraphics[width=\linewidth]{Figures/decision_tstat_talk.pdf}
%	\caption{Performance as measured by the normalized CFI for the Weighted Generalized Difference-in-Differences with $\gamma=0.6$ as a function of $t_c$ along with the 95\% confidence bands computed using cross-validation.} \label{fig:decision_tstat}
%	\endminipage
%\end{figure}

To interpret \cref{fig:decision_tstat_many} and \cref{fig:decision_tstat}, recall that for the policies of the form \cref{eq:decision_rule}, we roll out any treatments that achieve a $t$-statistic equal to or greater than $t_c$ for that ATE estimator. In other words, the Figures show the financial impact of requiring a low evidence threshold to roll out a treatment as $t_c \rightarrow -\infty$ and a large evidence threshold as $t_c \rightarrow \infty$. % progressively more evidence to roll a trusing a critical $t$-statistic $=\infty$ means not rolling out any treatment, while a critical $t$-statistic $=-\infty$ means rolling out all treatments irrespective of the evidence. 
For the specific case of Amazon Supply Chain, \cref{fig:decision_tstat} implies that the most significant improvement of the decision policy comes from adjusting $t$-critical to somewhere around $-1.2$ with a confidence interval of $(-2.4, 0.4)$. This change, and the roll outs that follow, could more than double the estimated cumulative financial impact of decisions made on the basis of our RCTs. %proposed treatments.

When it comes to the choice of estimator, the evidence is less convincing, at least for the three estimators shown in \cref{fig:decision_tstat_many}. We have not found the paired difference in normalized CFI for these estimators to be statistically significant when using $t_c=-1.2$. This is not too surprising given the width of the confidence bands in \cref{fig:decision_tstat}. 

% but the evidence here is much less convincing. Indeed the confidence \cref{fig:decision_tstat} shows the 95\% confidence intervals around $\hat{F}_E$ for the Weighted Generalized Difference-in-Differences relative to the current critical $t$-statistic of $1.96$. These confidence intervals are computed via cross-validation by re-sampling $(\cC_1, \cT_1)$ and $(\cC_2, \cT_2)$ a 100 times. As explained in \cite{bengio2004no}, confidence intervals computed via cross-validation should be interpreted cautiously.

\begin{figure}[!ht]
	\centering
		\hspace*{-0.75cm}\includegraphics[width=0.5\linewidth]{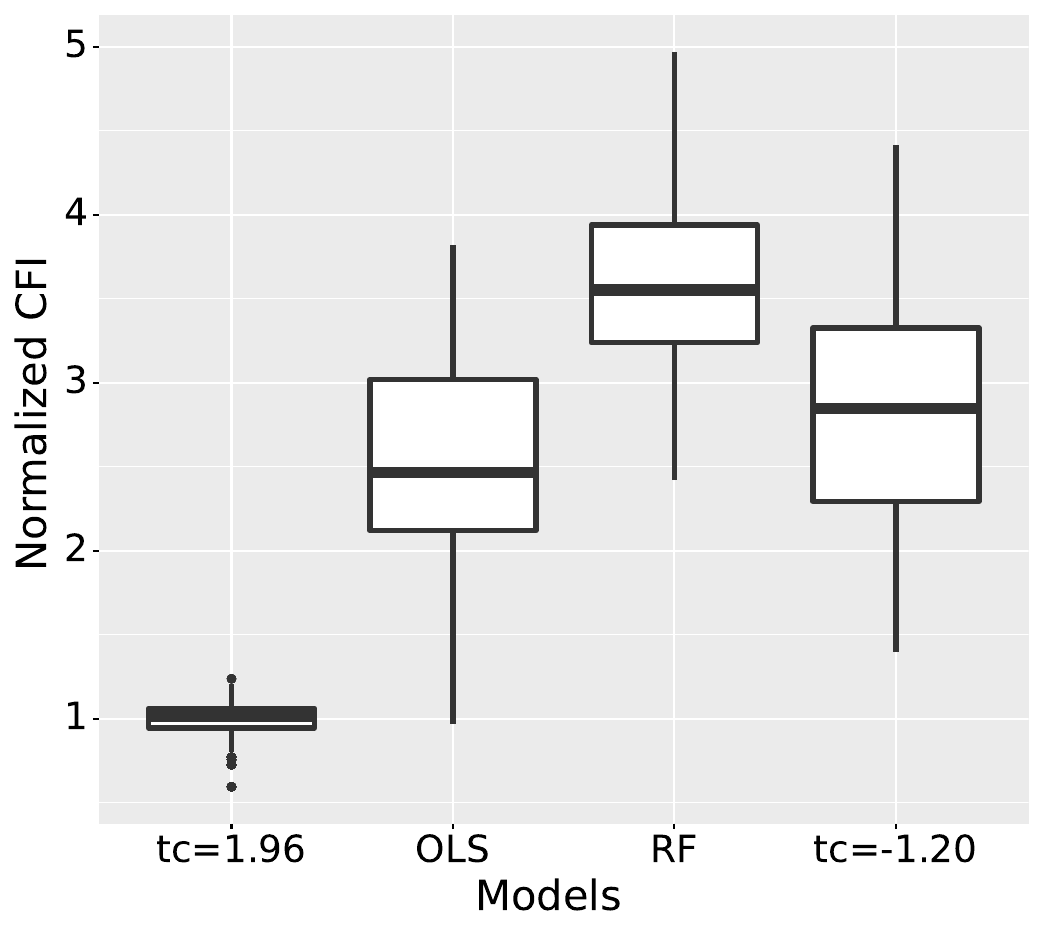}%)decision_alt_talk.pdf}
	%\subfloat[][]{\includegraphics[width=1\linewidth]{Figures/decision_alt.pdf}}
	\caption{Boxplot comparing the normalized CFI for 4 roll out policies: the standard policy with $t_c=1.96$; the Weighted Generalized Difference-in-Differences policy with $t_c=-1.2$; a linear regression (OLS) policy; and random forest (RF) based policies. The distributions are estimated using cross-validation.}
	\label{fig:decision_alt}

\end{figure}

Up until this point, we have limited our investigation to roll out policies of the form \cref{eq:decision_rule} based on a $t$-critical threshold for the ATE. \emph{A priori} there is no reason to expect that decision rules of this form will be the best choice to optimize \cref{eq:cum_ltfcf_hat} and so we also consider a more general framework and expand the policy space to regression models of the form $D = D(\X)$. Here the covariates $\X$ include pre-RCT variables, such as per unit profit and population size, but the  as well as different ATE estimates and their associated $t$-statistics $\frac{\hat{\Delta}_{E}}{\hat{\sigma}_{E}}$.
The plot in \cref{fig:decision_alt} shows that using a Random Forest for $D = D(\X)$  achieves the highest normalized CFI and this improvement over the second best model, the Weighted Generalized Difference-in-Differences with a $t$-critical of $-1.2$, is statistically significant.

\begin{figure}[!ht]
\centering
\! \!
\begin{minipage}[b]{0.45\linewidth}

\centering
	
\hspace*{-0.25cm}\includegraphics[width=1.1\linewidth]{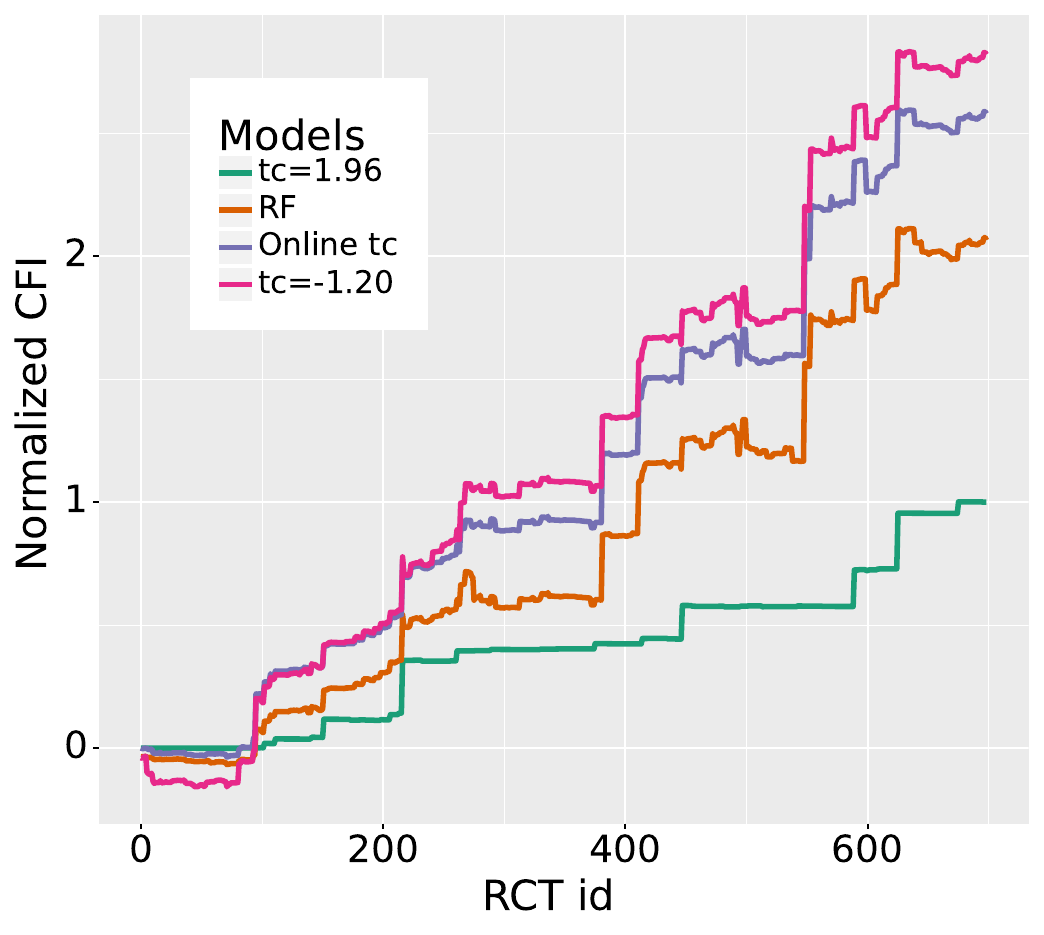}
\caption{Time evolution of the expected normalized CFI as an online decision problem where RCTs as ordered by end date. The plots compares four roll out policies: the  standard policy with $t_c=1.96$, an online random forest (RF); an online $t_c$ policy; and the Weighted Generalized Difference-in-Differences based policy using the optimal value of $t_c=-1.2$ .}
\label{fig:online_decision}

\end{minipage}
\quad \quad
\begin{minipage}[b]{0.45\linewidth}
\hspace*{-0.75cm}\includegraphics[width=1.1\linewidth]{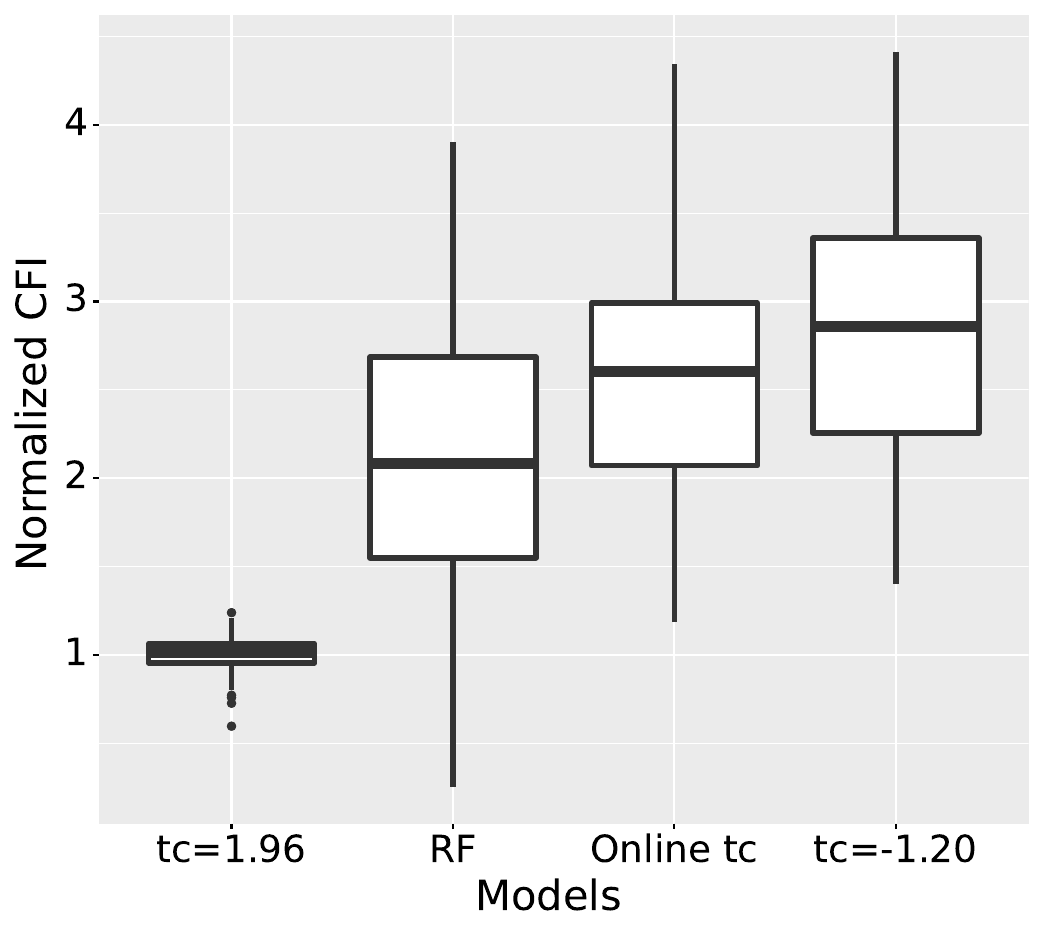}%)decision_alt_talk.pdf}
%\subfloat[][]{\includegraphics[width=1\linewidth]{Figures/decision_alt.pdf}}
\caption{Boxplot of the normalized CFI for the final state of the online decision problem for four policies: the  standard policy with $t_c=1.96$, an online random forest (RF); an online $t_c$ policy; and the Weighted Generalized Difference-in-Differences based policy using the optimal value of $t_c=-1.2$. The distributions are estimated using cross-validation.} %We also include an ``Oracle'' model for comparison which shows the maximum normalized CFI value that can be achieved in hindsight.}
\label{fig:online_decision_boxplot}
	
\end{minipage}
\quad
\end{figure}

 %the covariates $\X$ include pre-trial RCT variables, such as per unit profit and population size, but the  as well as the $t$-statistics $\frac{\hat{\Delta}_{E}}{\hat{\sigma}_{E}}$ from multiple ATE estimators. The question we are trying to answer is whether  The plot in \cref{fig:decision_alt} shows that in our case Random Forest achieves the highest normalized CFI, but this improvement over the Weighted Generalized Difference-in-Differences with a $t$-critical of $-1.2$ is not statistically significant. %As such it is unclear whether   using regression-based decision rules perform just as well as Weighted Generalized Difference-in-Differences with a critical $t$-statistic of $-1.2$.

%However, in our framework, there is no particular reason to do this. We extend our work to cover the space of roll out policies to more general regression models of the form $D = D(\X)$, where the covariates $\X$ include the t-statistic from multiple ATE estimators as well as some pre-trial RCT specific variables such as per unit profit and population size. \cref{fig:decision_alt} shows that using regression-based models performs just as well as Weighted Generalized Difference-in-Differences with a critical $t$-statistic of $-1.96$ performs.

%So in hindsight, i.e. test the roll out policy's performance on the same data used to train the model, a non-linear model using multiple ATE estimates, t-statistics, and pre-RCT variables out performs any roll out policy based purely on $t_c$. 
One lingering question however is how well this approach generalizes: that is, can a roll out policy learned on the 699 RCTs be safely used on future experiments? To answer this question, we revised the computation of \cref{eq:cum_ltfcf_hat} to capture the fact that the RCTs are not performed all at once, but rather they are ordered in time based on when each RCT ended. As such, optimizing $\hat{F}_E$ can be treated as an online regression problem where the outcome of the first $N$ RCTs should guide the policy (say regression coefficients) for the $N+1$st RCT. This both ensures the policy can adjust over time--which is important if the the ATE prior $\mathbb{P}(\Delta)$ experiences a distributional shift--but also that ensures that any claimed cumulative financial impact improvement is evaluated out-of-sample. 

\cref{fig:online_decision} shows the expected outcome of this procedure as RCTs accumulate for 4 policies: the standard policy from \cref{eq:standard_rule}, an online Random Forest (RF), an online $t_c$ roll out policy\footnote{That is to say the $t_c$ value used to decide RCT $N+1$ is based on the optimal value for the previous $N$ RCTs.} using Weighted General Difference-in-Differences, and the optimal $t_c=-1.2$ value for the Weighted General Difference-in-Difference. Not surprisingly, the optimal $t_c=-1.2$ in hindsight outperforms the online models. More important is the fact that both online models achieve higher normalized CFI than the standard policy \cref{eq:standard_rule} by a factor greater than two over the course of the RCT corpus, as shown in \cref{fig:online_decision_boxplot}. In other words, even without the benefit of hindsight applying either of these online policies would have lead to substantial improvement in CFI.

\section{Conclusion}

In this work, we develop a simple methodology for treatment effect model/estimator selection which pools the performance of estimators across RCTs. The methodology allows us to compare estimators on a held-out data fold in an unbiased way. The results align with a priori intuitions of estimator performance for our data corpus. One insight is that we should be trading off variance for more bias to reduce the MSE of treatment effect estimation in problems with heavy tails. 
% In particular, reweighting the pretreatment least-squares estimator and Winsorization both have the potential to improve the accuracy of treatment effect estimation. 
Further investigation into better estimators (as judged by their held-out MSE) and their coverage is warranted. 
% Similarly, investigating the theoretical properties and limitations of our aggregation framework is worthwhile.
The methodology also naturally lends itself to the question of when to roll out treatments, by allowing the comparison of different roll out policies based based on their estimated cumulative financial impact. We found that the standard policy of rolling out treatments for which the estimated ATE is significantly positive is far from optimal for our RCT corpus. In particular, a much more aggressive roll out policy can more than double the financial impact of decisions based on the RCTs run at Amazon.

While our corpus consists of RCTs at Amazon run over several years, we hope our primary methodological contribution -- to propose a cross-validation-like methodology to evaluate TE estimators and their corresponding decisions -- can be used to objectively evaluate causal inference techniques in settings where large corpora of RCTs are available.

	%!TEX root = main.tex
\section{Acknowledgements}

The authors thank Robert Stine, Edo Airoldi, and Kenny Shirley for their valuable comments and feedback.
	\newpage
	\appendix

	%!TEX root = main.tex
\section{Proofs of Estimator Validation Lemmas} \label{appendix:proofs}

First, we present the proof of \cref{claim:1}.

\begin{proof}[Proof of \cref{claim:1}]
	We simplify the MSE of a treatment effect estimator $E$ by centering the DM estimator around its mean and expanding the square:
	\begin{align}
		& \mE[(\hat{\Delta}_{A}(\cT_1, \cC_1) - \hat{\Delta}_{DM}(\cT_2, \cC_2))^2] = \mE[(\hat{\Delta}_{A}(\cT_1, \cC_1) - \Delta + \Delta - \hat{\Delta}_{DM}(\cT_2, \cC_2))^2] = \notag \\
		& \mE[(\hat{\Delta}_{A}(\cT_1, \cC_1) - \Delta)^2] + \mE[(\Delta - \hat{\Delta}_{DM}(\cT_2, \cC_2))^2] + 2 \mE[(\hat{\Delta}_{A}(\cT_1, \cC_1) - \Delta)] \cancelto{0}{\mE[(\Delta - \hat{\Delta}_{DM}(\cT_2, \cC_2))]} \implies \notag \\
		& \mE[(\hat{\Delta}_{A}(\cT_1, \cC_1) - \hat{\Delta}_{DM}(\cT_2, \cC_2))^2] = \mE[(\hat{\Delta}_{A}(\cT_1, \cC_1) - \Delta)^2] + \mE[(\Delta - \hat{\Delta}_{DM}(\cT_2, \cC_2))^2], \label{eq:pf3} 
	\end{align}
	where the cancellation uses the independence of the first/second folds of data to factor the expectation over the two terms, and the unbiased estimation property of the DM estimator over the second fold \citep{rubin2005causal}\footnote{Throughout we also implicitly use the fact the subfolds are (uniformly) randomly sampled from the treatment and control groups---so the expectation over the subfold is equivalent to the expectations over the entire treatment/control groups.}. We then obtain the following variances for two estimators $A$ and $B$:
	\begin{align}
		& \mE[(\hat{\Delta}_{A}(\cT_1, \cC_1) - \hat{\Delta}_{DM}(\cT_2, \cC_2))^2] - \mE[(\hat{\Delta}_{B}(\cT_1, \cC_1) - \hat{\Delta}_{DM}(\cT_2, \cC_2))^2] = \\ & \mE[(\hat{\Delta}_{A}(\cT_1, \cC_1) - \Delta)^2] - \mE[(\hat{\Delta}_{B}(\cT_1, \cC_1) - \Delta)^2], \label{eq:pffinal}
	\end{align}
	from which the claim follows.
\end{proof}

Next, we present the proof of \cref{claim:2}.
\begin{proof}[Proof of \cref{claim:2}]
	The proof is very much in keeping with \ref{claim:1} and relies on the unbiasedness of $\hat{\Delta}_{DM, i}$ and independence of the two splits $(\cT_1, \cC_1)$ and $(\cT_2, \cC_2)$:
	\begin{align}
		\mE[ \hat{F}_E]   & =  \mE \left [ \sum_{i\in\mathcal{I}} M_i \hat{\Delta}_{DM, i}(\cT_2, \cC_2) D_{E,i}(\cT_1, \cC_1) \right ] \notag = \sum_{i\in\mathcal{I}} M_i \mE[\hat{\Delta}_{DM, i}(\cT_2, \cC_2)]  \mE [ D_{E,i}(\cT_1, \cC_1)] \notag \\
		& = \sum_{i\in\mathcal{I}} M_i \Delta_i \mE [ D_{E,i}(\cT_1, \cC_1)], 
	\end{align}
	as claimed.
\end{proof}

\section{Additional Results}
\label{app:additional}
% All experiments herein were run on a large CPU instance (no GPUs were needed) on Amazon AWS with 144 cores and 2TB RAM. 

First we present several additional estimator histograms.

\begin{figure}[!htb]
	\centering
	\begin{minipage}[b]{0.3\linewidth}
		\centering
		\includegraphics[width=0.8\linewidth]{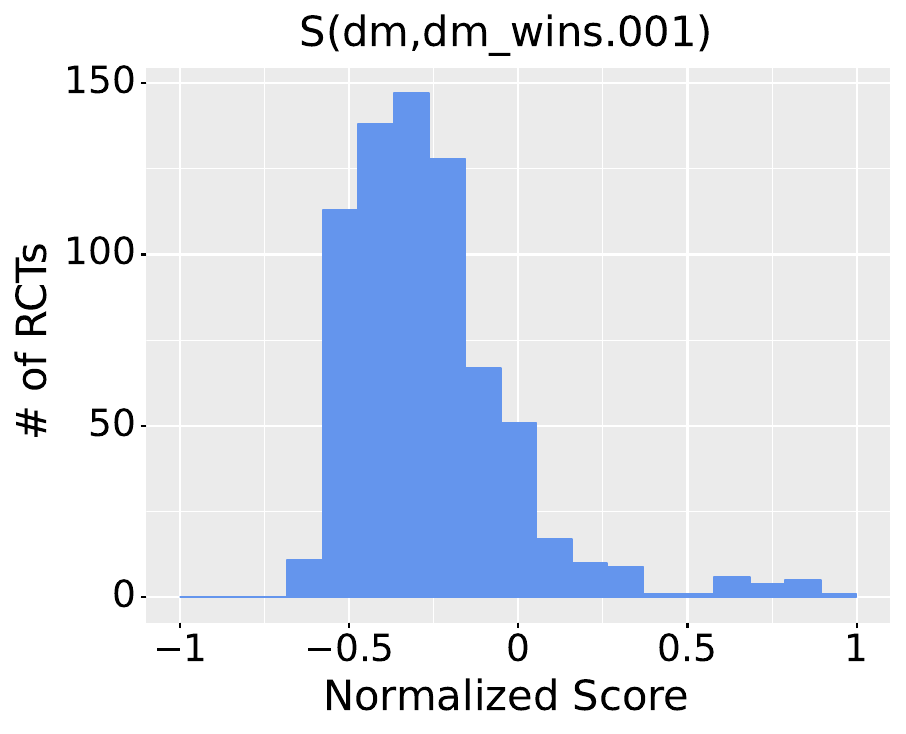}
		\caption{Histogram of the score distribution for dm vs Winsorized (at $0.001$) dm estimator.}
		\label{fig:score_hist_dm_dm_wins2}
	\end{minipage}
	\quad
	\begin{minipage}[b]{0.3\linewidth}
		\centering
		\includegraphics[width=0.8\linewidth]{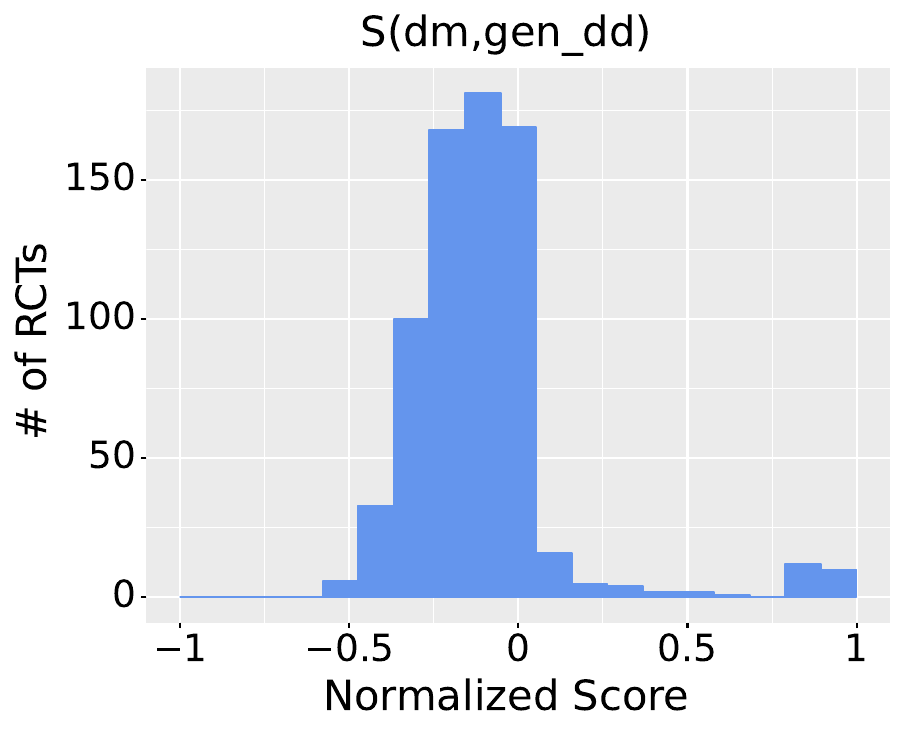}
		\caption{Histogram of the score distribution for dm vs gen$\_$dd estimator.}
		\label{fig:score_hist_dm_gen_dd2}
	\end{minipage}
	\quad
	\begin{minipage}[b]{0.3\linewidth}
		\centering
		\includegraphics[width=0.8\linewidth]{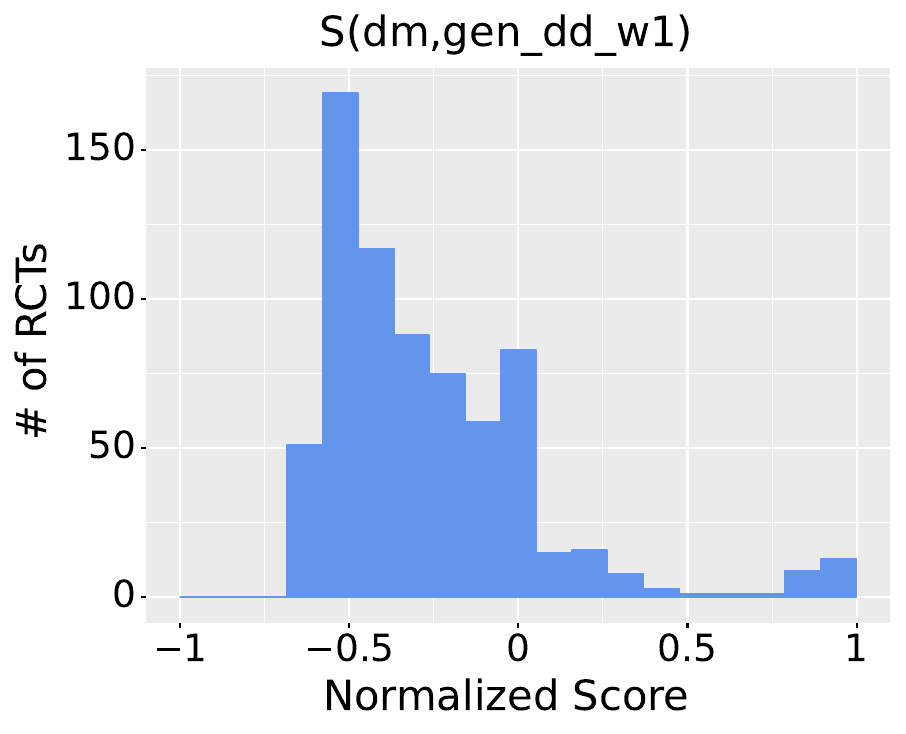}
		\caption{Histogram of the score distribution for dm vs gen$\_$dd$\_$w1 estimator.}
		\label{fig:score_hist_dm_gen_dd_w12}
	\end{minipage}
\end{figure}

\begin{figure}[!htb]
	\centering
	\begin{minipage}[b]{0.3\linewidth}
		\centering
		\includegraphics[width=0.8\linewidth]{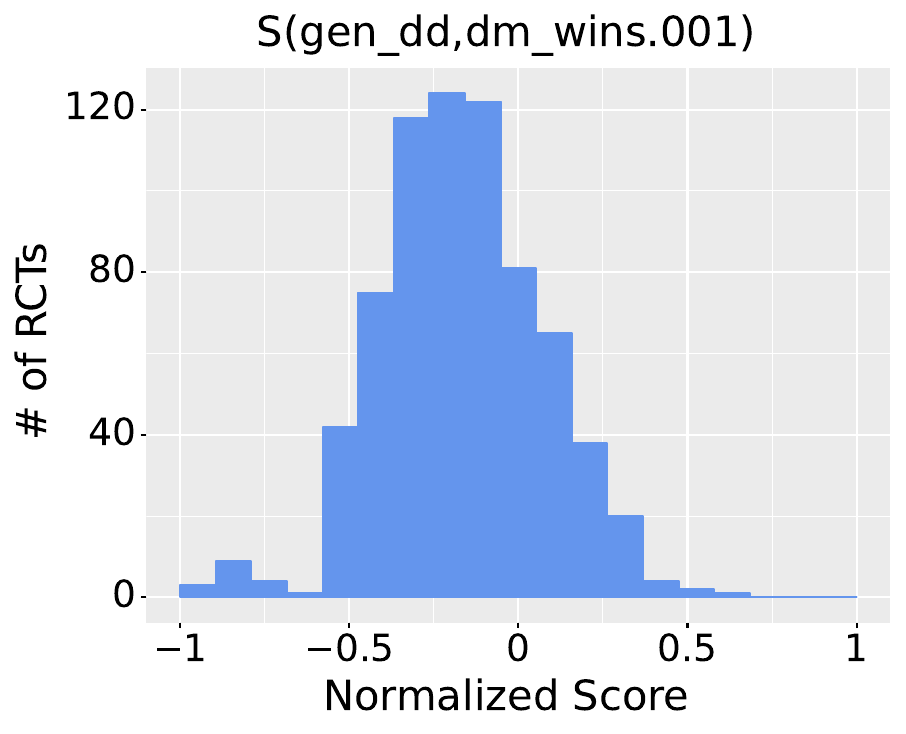}
		\caption{Histogram of the score distribution for gen$\_$dd vs Winsorized (at $0.001$) dm estimator.}
		\label{fig:score_hist_gen_dd_dm_wins}
	\end{minipage}
	\quad
	\begin{minipage}[b]{0.3\linewidth}
		\centering
		\includegraphics[width=0.8\linewidth]{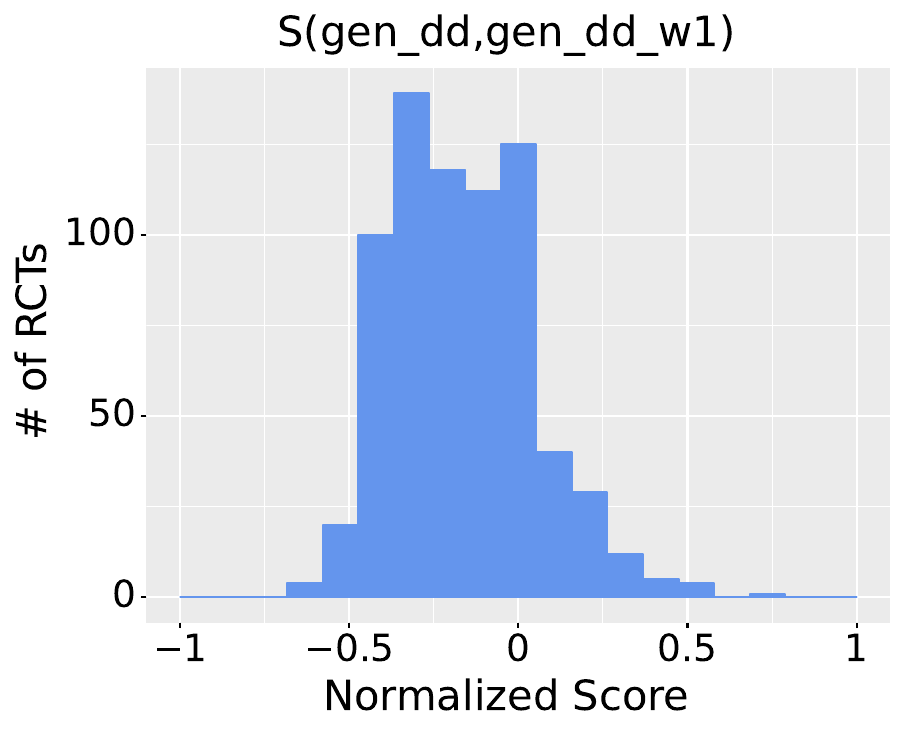}
		\caption{Histogram of the score distribution for gen$\_$dd vs gen$\_$dd$\_$w1 estimator.}
		\label{fig:score_hist_gen_dd_gen_dd_w1}
	\end{minipage}
	\quad
	\begin{minipage}[b]{0.3\linewidth}
		\centering
		\includegraphics[width=0.8\linewidth]{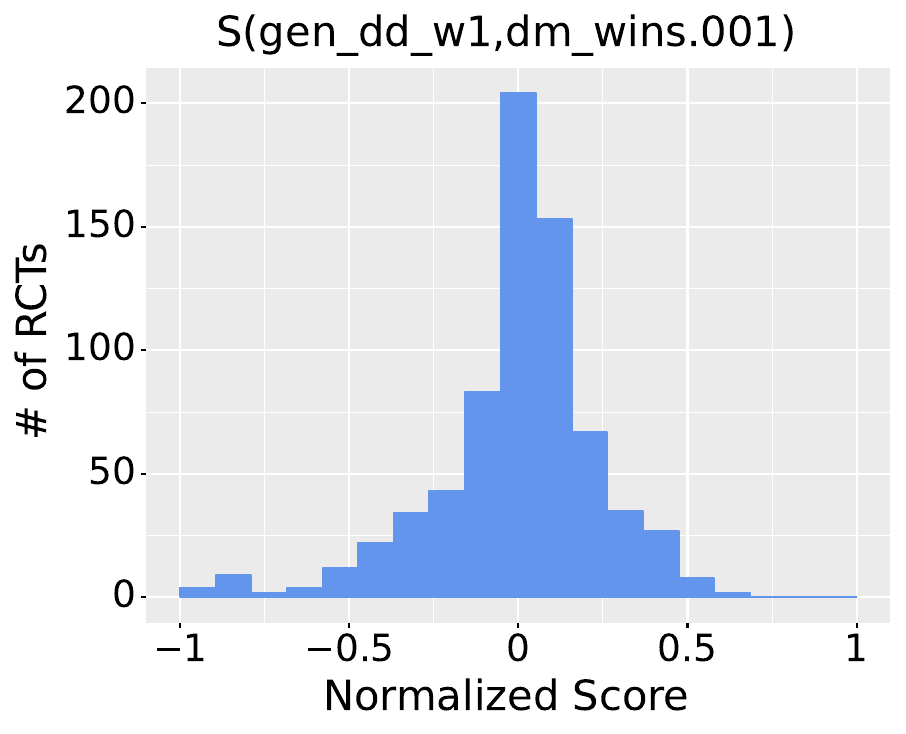}
		\caption{Histogram of the score distribution for gen$\_$dd$\_$w1 vs Winsorized (at $0.001$) dm estimator.}
		\label{fig:score_hist_gen_dd_w1_dm_wins}
	\end{minipage}
\end{figure}

\begin{figure}[!ht]
	\centering
	\begin{minipage}[b]{0.3\linewidth}
		\centering
		\includegraphics[width=0.85\linewidth]{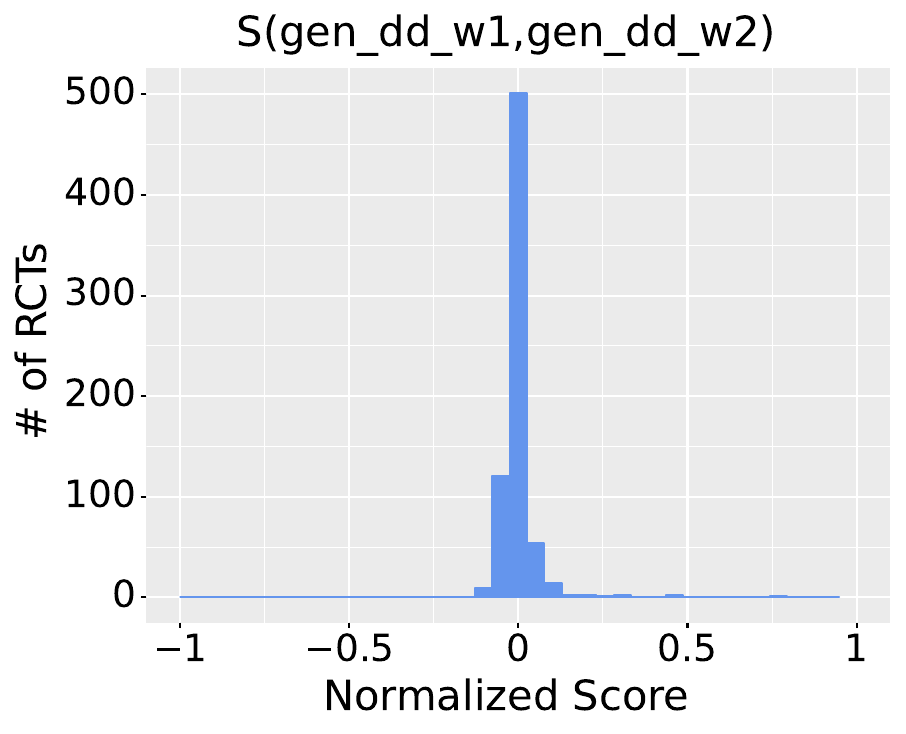}
		\caption{Histogram of the score distribution for gen$\_$dd$\_$w1 vs gen$\_$dd$\_$w2 estimator.}
		\label{fig:score_hist_gen_dd_w1_gen_dd_w3}
	\end{minipage}
	\quad
	\begin{minipage}[b]{0.3\linewidth}
		\centering
		\includegraphics[width=0.85\linewidth]{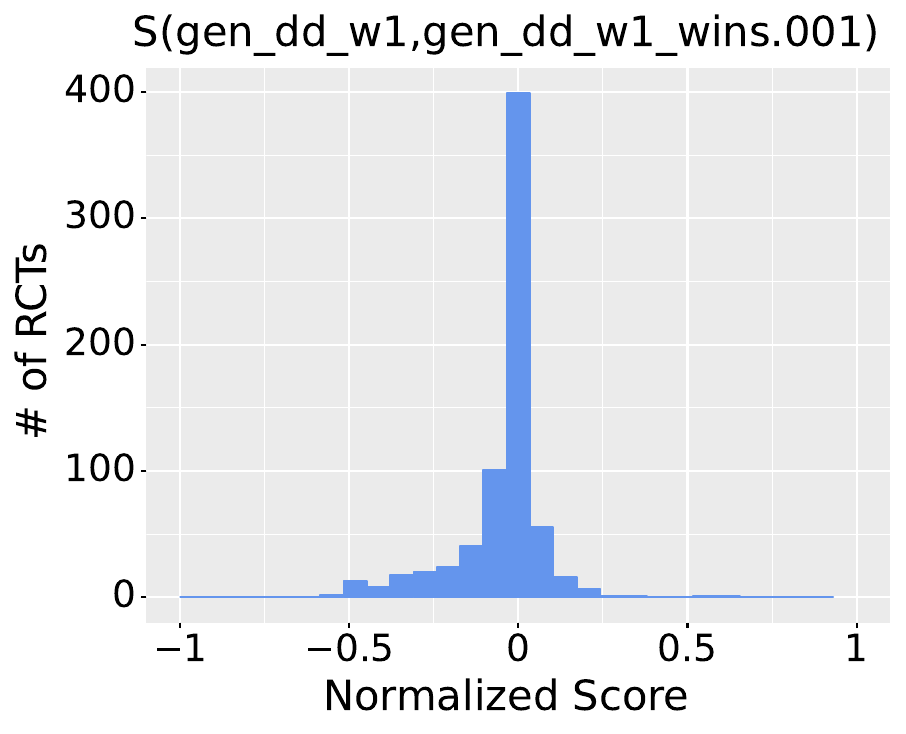}
		\caption{Histogram of the score distribution for gen$\_$dd$\_$w1 vs gen$\_$dd$\_$w1$\_$wins.001 estimator.}
		\label{fig:score_hist_gen_dd_w1_gen_dd_w1_wins}
	\end{minipage}
	\quad
	\begin{minipage}[b]{0.3\linewidth}
		\centering
		\includegraphics[width=0.85\linewidth]{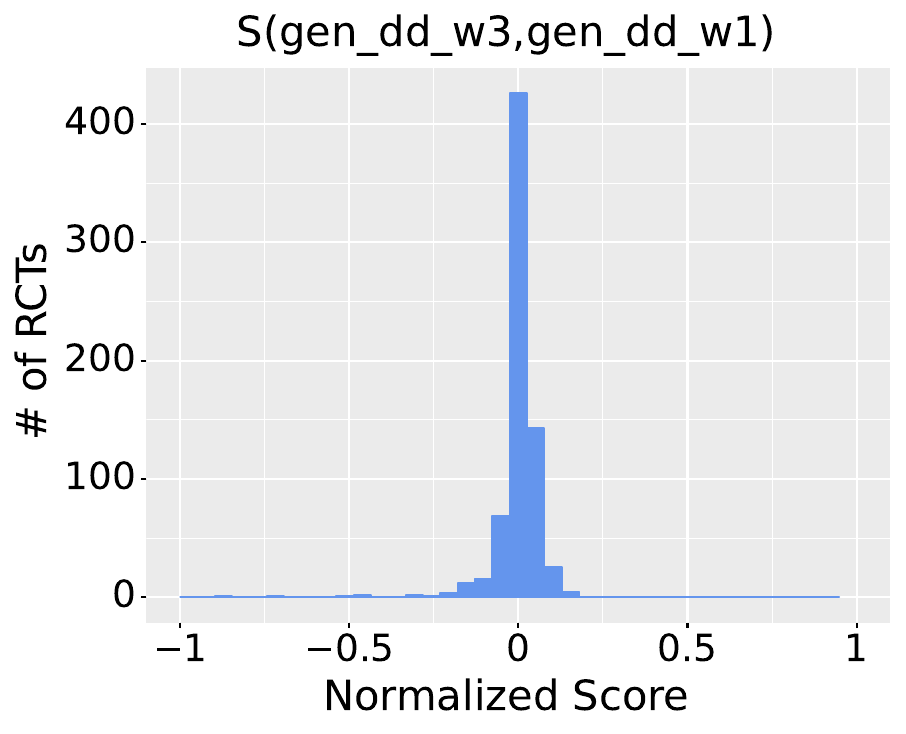}
		\caption{Histogram of the score distribution for gen$\_$dd$\_$w3 vs gen$\_$dd$\_$w1 estimator.}
		\label{fig:score_hist_mom1000}
	\end{minipage}
\end{figure}

In this section we present additional results from our aggregation methodology to explore their stability under using different bootstrapped train/test splits to compute the normalized score vectors $\hat{\A}$ and $\hat{\B}$. \cref{tab:tab1to50,,tab:tab51to100} show consistent results.

\begin{table}[!htb]
	\centering
	\resizebox{\columnwidth}{!}{%
		\begin{tabular}{|l|l|l|l|l|l|l|l|l|}
			\hline
			Method &
			dm &
			mom1000 &
			gen\_dd &
			gen\_dd\_w1 &
			gen\_dd\_w\_norm &
			dm\_wins.001 &
			gen\_dd\_wins.001 &
			gen\_dd\_w1\_wins.001 \\ \hline
			dm &
			x &
			(-3.58, 0.000369) &
			(-12.49, 1.68e-32) &
			(-21.95, 6.88e-82) &
			(-17.57, 1.27e-57) &
			(-27.53, 5.04e-114) &
			(-24.74, 6.91e-98) &
			(-24.47, 2.51e-96) \\ \hline
			mom1000 &
			(3.58, 0.000369) &
			x &
			(-2.03, 0.043) &
			(-11.7, 5.02e-29) &
			(-9.15, 5.86e-19) &
			(-13.23, 7.35e-36) &
			(-14.3, 6.26e-41) &
			(-15.43, 1.68e-46) \\ \hline
			gen\_dd &
			(12.49, 1.68e-32) &
			(2.03, 0.043) &
			x &
			(-20.42, 3.05e-73) &
			(-13.23, 7.11e-36) &
			(-18.44, 2.49e-62) &
			(-24.2, 9.27e-95) &
			(-22.75, 1.9e-86) \\ \hline
			gen\_dd\_w1 &
			(21.95, 6.88e-82) &
			(11.7, 5.02e-29) &
			(20.42, 3.05e-73) &
			x &
			(6.83, 1.8e-11) &
			(-0.22, 0.828) &
			(-4.82, 1.78e-06) &
			(-9.39, 7.72e-20) \\ \hline
			gen\_dd\_w\_norm &
			(17.57, 1.27e-57) &
			(9.15, 5.86e-19) &
			(13.23, 7.11e-36) &
			(-6.83, 1.8e-11) &
			x &
			(-4.38, 1.37e-05) &
			(-8.76, 1.46e-17) &
			(-11.22, 5.47e-27) \\ \hline
			dm\_wins.001 &
			(27.53, 5.04e-114) &
			(13.23, 7.35e-36) &
			(18.44, 2.49e-62) &
			(0.22, 0.828) &
			(4.38, 1.37e-05) &
			x &
			(-4.03, 6.21e-05) &
			(-5.27, 1.79e-07) \\ \hline
			gen\_dd\_wins.001 &
			(24.74, 6.91e-98) &
			(14.3, 6.26e-41) &
			(24.2, 9.27e-95) &
			(4.82, 1.78e-06) &
			(8.76, 1.46e-17) &
			(4.03, 6.21e-05) &
			x &
			(-4.11, 4.44e-05) \\ \hline
			gen\_dd\_w1\_wins.001 &
			(24.47, 2.51e-96) &
			(15.43, 1.68e-46) &
			(22.75, 1.9e-86) &
			(9.39, 7.72e-20) &
			(11.22, 5.47e-27) &
			(5.27, 1.79e-07) &
			(4.11, 4.44e-05) &
			x \\ \hline
		\end{tabular}%
	}
	\caption{Comparison of Estimators via one-sample $t$-test applied to their normalized score vector. This table was computed using error vectors from only 50 resampled train/test splits to feed into $\hat{\A}$ and $\hat{\B}$. Easiest to read row-wise. The index $(A, B)$ of the table computes the pair of the ($t$-statistic, $p$-value) associated with the score $\score(\hat{\A}, \hat{\B})$. A large positive $t$-statistic at index $(A, B)$ indicates estimator $A$ is better then estimator $B$ and vice versa.}
	\label{tab:tab1to50}
\end{table}

\begin{table}[!htb]
	\centering
	\resizebox{\columnwidth}{!}{%
		\begin{tabular}{|l|l|l|l|l|l|l|l|l|}
			\hline
			Method &
			dm &
			mom1000 &
			gen\_dd &
			gen\_dd\_w1 &
			gen\_dd\_w\_norm &
			dm\_wins.001 &
			gen\_dd\_wins.001 &
			gen\_dd\_w1\_wins.001 \\ \hline
			dm &
			x &
			(-3.44, 0.000613) &
			(-12.49, 1.73e-32) &
			(-22.14, 5.73e-83) &
			(-17.7, 2.55e-58) &
			(-27.85, 7.64e-116) &
			(-25.18, 2.11e-100) &
			(-24.73, 8.29e-98) \\ \hline
			mom1000 &
			(3.44, 0.000613) &
			x &
			(-2.24, 0.0252) &
			(-11.93, 5.02e-30) &
			(-9.39, 8.29e-20) &
			(-13.64, 8.54e-38) &
			(-14.76, 3.62e-43) &
			(-15.8, 2.07e-48) \\ \hline
			gen\_dd &
			(12.49, 1.73e-32) &
			(2.24, 0.0252) &
			x &
			(-20.86, 1.08e-75) &
			(-13.48, 4.75e-37) &
			(-18.83, 2e-64) &
			(-24.98, 2.95e-99) &
			(-23.22, 4.23e-89) \\ \hline
			gen\_dd\_w1 &
			(22.14, 5.73e-83) &
			(11.93, 5.02e-30) &
			(20.86, 1.08e-75) &
			x &
			(6.72, 3.64e-11) &
			(-0.37, 0.714) &
			(-5.31, 1.47e-07) &
			(-9.42, 6.27e-20) \\ \hline
			gen\_dd\_w\_norm &
			(17.7, 2.55e-58) &
			(9.39, 8.29e-20) &
			(13.48, 4.75e-37) &
			(-6.72, 3.64e-11) &
			x &
			(-4.52, 7.21e-06) &
			(-9.1, 8.98e-19) &
			(-11.23, 5e-27) \\ \hline
			dm\_wins.001 &
			(27.85, 7.64e-116) &
			(13.64, 8.54e-38) &
			(18.83, 2e-64) &
			(0.37, 0.714) &
			(4.52, 7.21e-06) &
			x &
			(-4.2, 3.05e-05) &
			(-5.32, 1.37e-07) \\ \hline
			gen\_dd\_wins.001 &
			(25.18, 2.11e-100) &
			(14.76, 3.62e-43) &
			(24.98, 2.95e-99) &
			(5.31, 1.47e-07) &
			(9.1, 8.98e-19) &
			(4.2, 3.05e-05) &
			x &
			(-3.87, 0.000119) \\ \hline
			gen\_dd\_w1\_wins.001 &
			(24.73, 8.29e-98) &
			(15.8, 2.07e-48) &
			(23.22, 4.23e-89) &
			(9.42, 6.27e-20) &
			(11.23, 5e-27) &
			(5.32, 1.37e-07) &
			(3.87, 0.000119) &
			x \\ \hline
		\end{tabular}%
	}
	\caption{Comparison of Estimators via one-sample $t$-test applied to their normalized score vector. This table was computed using error vectors from only 50 resampled train/test splits to feed into $\hat{\A}$ and $\hat{\B}$ distinct from those in previous table. Easiest to read row-wise. The index $(A, B)$ of the table computes the pair of the ($t$-statistic, $p$-value) associated with the score $\score(\hat{\A}, \hat{\B})$. A large positive $t$-statistic at index $(A, B)$ indicates estimator $A$ is better then estimator $B$ and vice versa.}
	\label{tab:tab51to100}
\end{table}

\section{Cross-Validation Methodology} \label{appendix:method}
The cross-validation methodologies described in \cref{sec:val} and \cref{sec:decision} are for the most part intuitive; nonetheless, it is worthwhile to present all the details of how we partition $\cT$ and $\cC$ as well as how we repeat the procedure to cross-validate \cref{eq:held_out} and \cref{eq:cum_ltfcf_hat}. We start with a formal definition of treatment $\cT$ and control $\cC$ groups. Let $i$ be some lab in $\mathcal{I}$; then, the treatment group for this lab is the set of outcomes $Y_{i, a}$ and features $X_{i, a}$ for each product $a$ under the in the treatment arm $T_{i,a} =1$, that is $\cT_i = \{(Y_{i, a}, X_{i, a})|T_{i, a}=1 \}$ where $|\cT_i| = K_i$ is the number of products that were assigned to the treatment arm $T_{i, a}=1$. Similarly, the control group is given by $\cC = \{(Y_{i, a}, X_{i, a})|T_{i, a} = 0 \}_{a=K_i+1}^{M_i}$ where $|\cC_i| = M_i-K_i$ is the number of products assigned to the control arm $T_{i, a}=0$ and $M_i$ is the total number of products in the lab. 

The goal of our methodology is to find an optimal estimator $\hat{\Delta}_i$ for the ATE $\Delta_i$ or an optimal roll out policy $D_i$ under some objective function $L$. This means finding a function of $\cT$ and $\cC$ such that $f(\cT, \cC) \in \mathbb{R}$ or $f(\cT, \cC) \in \{0, 1\}$ respectively, and that optimizes the expected objective:
\begin{align} \label{eq:abs_obj}
	\mE[L(f(\cT, \cC), \Delta)]
\end{align}
for $f$ in some functional space $\mathcal{F}$. As discussed in \cref{sec:val} and \cref{sec:decision}, to do this in the context of an RCT where $\Delta$ is unknown, we rely on the fact that the difference-in-means estimator $\hat{\Delta}(\cT, \cC)$ is unbiased for the ATE $\Delta$. Specifically, for any lab $i$, we randomly split the treatment and control group using two random subsets of product indices $S_{i} = \{1, \dots, K_i\}$ and $R_{i} = \{K_i+1, \dots, M_i\}$ so that we end up with the four following sets:
\begin{itemize}
	\item $\cT_{i, 1} = \{(Y_{i, a}, X_{i, a})|T_{i, a}=1 \,\, \text{and} \,\, a \in S_{i} \}$ 
	\item $\cC_{i, 1} = \{(Y_{i, a}, X_{i, a})|T_{i, a}=0 \,\, \text{and} \,\, a \in R_{i} \}$ 
	\item $\cT_{i, 2} = \{(Y_{i, a}, X_{i, a})|T_{i, a}=1 \,\, \text{and} \,\, a \notin S_{i} \}$ 
	\item $\cC_{i, 2} = \{(Y_{i, a}, X_{i, a})|T_{i, a}=0 \,\, \text{and} \,\, a \notin R_{i} \}$ .
\end{itemize}
We also pick the size of $S_{i}$ and $R_{i}$ so that the split proportion $p$ is constant across treatment, control, and labs:
\begin{equation*}
	\frac{|S_{i}|}{K_i} = \frac{|R_{i}|}{M_i - K_i} = p.
\end{equation*}

With this splitting methodology, we can now replace \cref{eq:abs_obj} with the empirical mean of the objective over all the labs in $\mathcal{I}$:
\begin{equation*}
	\frac{1}{|\mathcal{I}|}\sum_{i\in \mathcal{I}}L(f(\cT_{i, 1}, \cC_{i, 1}), \hat{\Delta}_{DM}(\cT_{i, 2}, \cC_{i, 2})).
\end{equation*}
We can now optimize empirical objective for $f$ similarly to empirical risk minimization for supervised learning. %However, we can also take advantage of the fact that there are multiple possible splits of $\cT$ and $\cC$. 
We can also ``cross-validate'' the empirical mean of the objective to reduce the subsampling variance and to get confidence intervals, as in \cref{fig:decision_tstat}. To do this we simply repeat the splitting procedure multiple times so that every random index set $S_{i}$ and $R_{i}$ is now also indexed by a split $b\in\{1, \dots, B\}$. Putting all of this together, we now have:
\begin{equation} \label{eq:emp_obj}
	\frac{1}{|\mathcal{I}|}\sum_{i\in \mathcal{I}}\frac{1}{B}\sum_{b=1}^{B}L(f(\cT_{i, b, 1}, \cC_{i, b, 1}), \hat{\Delta}_{DM}(\cT_{i, b, 2}, \cC_{i, b, 2})).
\end{equation}
This is how we estimated \cref{eq:held_out} and \cref{eq:cum_ltfcf_hat} in the paper, using $p=0.5$ and $B=100$. It is worth noting that in the case of \cref{eq:held_out}, we ended up replacing the outer sum of \cref{eq:emp_obj} with the aggregation methodology of \cref{sec:agg} to deal with the heavy-tailed nature of $M_i$, i.e. to ensure that the largest labs did not dominate the value of \cref{eq:emp_obj}.

	%\acks{The authors thank Robert Stine, Edo Airoldi, and Kenny Shirley for their valuable comments and feedback. All work was performed while working at/for Amazon.}
	
	% Manual newpage inserted to improve layout of sample file - not
	% needed in general before appendices/bibliography.

	% Note: in this sample, the section number is hard-coded in. Following
	% proper LaTeX conventions, it should properly be coded as a reference:
	
	%In this appendix we prove the following theorem from
	%Section~\ref{sec:textree-generalization}:

	\vskip 0.2in
	\bibliography{ref}

\end{document}